\documentclass[breaklinks, onecolumn, colorlinks, mathlines]{aastex63}   
\hypersetup{linkcolor=blue,citecolor=blue,filecolor=cyan,urlcolor=magenta}

\usepackage{geometry}
\usepackage{xcolor}
\usepackage{graphicx}
\usepackage{amssymb}
\usepackage{amsmath}
\usepackage{epstopdf}
\usepackage{mathrsfs}
\usepackage{revsymb}
\usepackage{color}

\def\be{\begin{equation}}
\def\ee{\end{equation}}

\def\teq#1{$\, #1\,$}                         

\def\teq#1{$\, #1\,$}                         
\def\pmb#1{\setbox0=\hbox{#1}%
  \kern-0.0125em\copy0\kern-\wd0
  \kern0.025em\copy0\kern-\wd0
  \kern-0.0125em\raise0.0433em\box0 }
\def\boxit#1{\vbox{\hrule\hbox{\vrule\kern3pt
       \vbox{\kern3pt #1\kern3pt}\kern3pt\vrule}\hrule}}
\def\mboxit#1{\vbox{\hrule\hbox{\vrule\kern6pt
       \vbox{\kern6pt \hbox{$\displaystyle #1$} 
             \kern6pt}\kern6pt\vrule}\hrule}}

\def\dover#1#2{\hbox{${{\displaystyle#1 \vphantom{(} }\over{
\displaystyle #2 \vphantom{(} }}$}}
{\catcode`\@=11                                                  
\gdef\SchlangeUnter#1#2{\lower2pt\vbox{\baselineskip 0pt\lineskip0pt    
\ialign{$\m@th#1\hfil##\hfil$\crcr#2\crcr\sim\crcr}}}}           
\def\gtrsim{\mathrel{\mathpalette\SchlangeUnter>}}               
\def\lesssim{\mathrel{\mathpalette\SchlangeUnter<}}    
\font\fiverm=cmr5

          \font\sixrm=cmr6

\def\sigt{\sigma_{\hbox{\sixrm T}}}
\def\taut{\tau_{\hbox{\sixrm T}}}

\def\timeIU{t_{\hbox{\sixrm IU}}}
\def\timeAP{t_{\hbox{\sixrm AP}}}
\def\taueff{\tau_{{\hbox{\sixrm eff}}}}
\def\wcyc{\omega_{\hbox{\fiverm B}}}
\def\thetaB{\theta_{\hbox{\fiverm B}}}
\def\kvec{\boldsymbol{k}}
\def\kvechat{\hat{\boldsymbol{k}}}
\def\Bvec{\boldsymbol{B}}
\def\Bvechat{\hat{\boldsymbol{B}}}

\def\calEvec{\boldsymbol{\cal E}}

\def\calEthet{{\cal E}_\theta}
\def\calEphi{{\cal E}_\phi}

\def\SigmaB{\Sigma_{\hbox{\sixrm B}}}
\def\DeltaB{\Delta_{\hbox{\sixrm B}}}

\def\betaparsq{\beta^2_{\parallel\hbox{\sixrm B}}}
\def\betaperpsq{\beta^2_{\perp\hbox{\sixrm B}}}
\def\const#1#2{\hbox to \hsize{\teq{#1} #2\hfill}}

\usepackage{lineno}
\usepackage{ulem}
\newcommand{\ored}[1]{{\color{red}\sout{#1}}}

\newcommand{\stkout}[1]{\ifmmode\text{\sout{\ensuremath{\red{ #1 }}}}\else\ored{#1}\fi}

\usepackage{empheq}
\usepackage[most]{tcolorbox}

\begin{document}

\title{\uppercase{Magnetic Thomson Transport in High Opacity Domains}}

\correspondingauthor{Matthew G. Baring}
\email{baring@rice.edu}

\author[0000-0003-4433-1365]{Matthew G. Baring}
\affiliation{Department of Physics and Astronomy - MS 108, Rice University, 
   6100 Main Street, Houston, Texas, 77251-1892, USA}
\author[0000-0002-9705-7948]{Kun Hu}
\affiliation{Physics Department, McDonnell Center for the Space Sciences, 
   Washington University in St. Louis, St. Louis, MO 63130, USA}
\author[0000-0001-9268-5577]{Hoa Dinh Thi}
\affiliation{Department of Physics and Astronomy - MS 108, Rice University, 
   6100 Main Street, Houston, Texas, 77251-1892, USA}

\begin{abstract}
X-ray radiation from neutron stars manifests itself in a variety of settings.
Isolated pulsars, and magnetars both exhibit quasi-thermal persistent soft X-ray
emission from their surfaces. Transient magnetospheric bursts from magnetars and
pulsed signals from accreting neutron stars mostly appear in harder X rays. The
emission zones pertinent to these signals are all highly Thomson optically thick
so that their radiation anisotropy and polarization can be modeled using
sophisticated simulations of scattering transport from extended emission
regions.  Validation of such codes and their efficient construction is enhanced
by a deep understanding of scattering transport in high opacity domains.  This
paper presents a new analysis of the polarized magnetic Thomson radiative
transfer in the asymptotic limit of high opacity.  The integro-differential
equations for photon scattering transport that result from a phase matrix
construction are reduced to a compact pair of equations.  This pair is then
solved numerically for two key parameters that describe the photon anisotropy
and polarization configuration of high Thomson opacity environs. Empirical
approximations for these parameters as functions of the ratio of the photon and
cyclotron frequencies are presented. Implementation of these semi-analytic
transport solutions as interior boundary conditions is shown to speed up
scattering simulations.  The solutions also enable the specification of the
anisotropic radiation pressure.  The analysis is directly applicable to the
atmospheres of magnetars and moderate-field pulsars, and to the accretion
columns of magnetized X-ray binaries, and can be adapted to address other
neutron star settings.
\end{abstract}



\section{Introduction}
 \label{sec:Intro}
Compton scattering has long been invoked in a variety of contexts for
neutron star emission, spanning their pulsed surface X-ray luminosity to their persistent 
and transient hard X-ray signals emanating from their magnetospheres.  
Central to considerations in the emission zones is the anisotropy imposed by the 
strong magnetic fields of neutron stars.  These fields are typically around a TeraGauss, 
yet span a broad range from around 1 GigaGauss for millisecond pulsars 
(see the ATNF Pulsar Catalogue\footnote{https://www.atnf.csiro.au/research/pulsar/psrcat/}) 
to 1 PetaGauss for magnetars (see the McGill Magnetar Catalog \citep{Olausen-2014-ApJS} 
and its online portal\footnote{http://www.physics.mcgill.ca/~pulsar/magnetar/main.html}),
naturally placing the electron cyclotron frequency \teq{\wcyc} in the X-ray or gamma-ray band. 
In these domains, the resonance in the cross section that naturally arises 
at the cyclotron frequency \citep{Canuto-1971-PhRvD,Herold-1979-PhRvD} and its harmonics 
\citep{DH-1986-ApJ} strongly impacts the character of scattering anisotropy and polarization levels 
in optically thick environs.  

Principal astrophysical environments where such cyclotron resonance influences
are important include accreting 
X-ray binaries such as Her X-1, 4U 0115+63, Vela X-1 and AO 0535+26.
Cyclotron absorption features have been observed above around 10 keV 
in about three dozen systems \citep[see the review by][]{Staubert-2019-AandA},
and are used to calibrate their magnetic fields to the \teq{\gtrsim 10^{12}}Gauss domain
\citep[e.g.,][]{Truemper-1978-ApJ,Coburn-2002-ApJ}.
The leading interpretation is that these features are generated through scattering and 
absorption at the cyclotron resonances in their accretion columns at or just above the 
stellar surface.  The line detections have led to sophisticated codes modeling radiative transfer
and the accompanying formation of spectral lines
\citep[e.g.,][]{Isenberg-1998-ApJ,Araya-1999-ApJ,Schoenherr-2007-AandA,Schwarm-2017-AandA}.
These Thomson optically thick columns are funnels of infalling material collimated by the magnetic 
field lines above the polar surface regions.

Surface signals from isolated pulsars and magnetic white dwarfs are also subject to strong
cyclotronic influences on radiation emanating from their high opacity atmospheres.
In the case of white dwarfs, the character of anisotropy and polarization due to 
atmospheric magnetic Thomson scattering was explored
using Monte Carlo simulations by \cite{Whitney-1991-ApJS}.  Yet the white dwarf context 
is complicated by the presence of numerous atomic lines and mixtures in the atmospheric 
composition: see \cite{Wickramasinghe-2000-PASP} for a review, and 
\cite{Jewett-2024-ApJ} for a Gaia-era spectroscopic update.
The modeling of surface X-ray emission from neutron stars 
has received extensive treatment over the years.  Most studies 
use scattering and free-free opacity to mediate the photon transport and also help support the atmospheres
hydrostatically.  Older investigations 
focused on traditional neutron stars with surface polar fields \teq{B_p < 10^{13}}Gauss, 
mostly fully-ionized hydrogen/helium atmospheres at temperatures \teq{T\sim 10^6}K 
\citep{Shibanov-1992-AandA,Pavlov-1994-AandA,Zavlin-1996-AandA,Zane-2000-ApJ}.

A more recent focus has been on atmospheric emission from magnetars.  
This has primarily been because they are hot and therefore bright, and so are 
easily observable at distance of 10 kpc or more.  Their surface polar field strengths 
are typically in the range of \teq{10^{14} - 2 \times 10^{15}}Gauss, placing the 
electron cyclotron frequency \teq{\wcyc} in the gamma-ray band.  Accordingly, there is a strong 
suppression of the scattering cross section in the X-ray band \citep{Canuto-1971-PhRvD,Herold-1979-PhRvD},
particularly for one of the polarization eigenmodes.  This effect critically influences the
scattering process and emergent polarization from magnetar atmospheres, and has 
been explored at length using solutions of the radiative transfer equation by various groups 
\citep{Ho-Lai-2001-MNRAS,Ozel-2001-ApJ,Ho-Lai-2003-MNRAS,Ozel-2003-ApJ,Adelsberg-2006-MNRAS,Taverna-2015-MNRAS}, 
mostly using semi-analytic, Feautrier-type techniques.  Their analyses
included detailed treatment of the polarization of the vacuum by the strong magnetic field 
\citep{Tsai-1975-PhRvD}, an influence of quantum electrodynamics (QED) that renders the vacuum birefringent so that 
radiative transfer is sensitive to the polarization state (mode), just as it is for a plasma.  
There have also been Monte Carlo simulations of atmospheric radiative transport 
\citep{Bulik-1997-MNRAS,Niemiec-2006-ApJ,Barchas-2021-MNRAS,Hu-2022-ApJ} that 
enable great facility in treating various surface locales possessing different {\bf B} 
orientations to the zenith direction.

Magnetars also exhibit sporadic and highly super-Eddington 
burst emission, so luminous that their radiating plasma densities are optically thick to Compton scattering 
\citep[e.g.,][]{Taverna-2017-MNRAS}.  This leads to the appearance of broad,
non-thermal ``Comptonized''
hard X-ray spectra \citep[e.g.][]{Lin-2011-ApJ,Lin-2012-ApJ,Younes-2014-ApJ}.
Consequently, the intense radiation pressure in magnetar bursts (and even more so in the much rarer 
giant flares) must push the emitting plasma so that it moves relativistically within the burst emission regions.  
The speeds of the motions are governed by radiation-driven dynamics, which will influence 
the spectrum of the radiation we observe \citep{Lin-2012-ApJ,Roberts-2021-Nature},
and undoubtedly its polarization also.  Therefore, understanding the dependence 
of (polarized) radiation pressure on anisotropies induced by Compton scattering in the strong 
magnetic field is an important element of constructing realistic emission models 
for magnetar bursts and giant flares.

A deeper appreciation of magnetic scattering transport in high opacity domains is 
therefore warranted, both to enhance the definition of boundary conditions for 
radiative transfer computations and transport simulations of emission
regions that are highly magnetized, and also to improve the understanding of 
the associated pressure anisotropy of the radiation field.   This is most conveniently 
delivered in the regime of magnetic Thomson scattering, for which the cross section 
is mathematically simple, and is not imbued with the complexities of QED modifications,
which include a multitude of resonant cyclotron harmonics.
Yet the magnetic Thomson domain is quite broadly applicable to white dwarfs 
in general, and to soft X-ray surface emission below 5 keV from neutron stars 
of different varieties, ranging from millisecond pulsars to Crab-like pulsars to magnetars; 
QED influences in strong fields \citep[e.g.][]{DH-1986-ApJ} generally only 
become important for photon energies \teq{\gtrsim 20}keV.

This paper presents a new analysis of the polarized magnetic Thomson 
radiative transfer in the asymptotic limit of high opacity.  The approach obtains 
solutions of the polarized radiation transport integro-differential equations, 
leveraging the simplicity of the magnetic Thomson differential cross section 
\citep{Canuto-1971-PhRvD,Herold-1979-PhRvD}.  In Sec.~\ref{sec:formalism}, 
the system is distilled down to two integral equations that constitute a transcendental 
system.   These are then solved numerically in Sec.~\ref{sec:solutions} for the two key 
parameters that describe the photon anisotropy and polarization configuration of high 
opacity environs.  After isolating analytic solutions in low and high frequency 
domains in Sec.~\ref{sec:special}, empirical approximations for these two parameters 
as functions of the ratio of the photon and cyclotron frequencies are developed in 
Sec.~\ref{sec:numerics}.  The Stokes parameter variations and resultant pressure 
anisotropies are addressed in Sec.~\ref{sec:anis_pol_char}.  Implementation of the 
semi-analytic transport solutions as interior boundary conditions in atmospheric slabs 
is shown in Sec.~\ref{sec:discuss} to speed up the magnetic Thomson scattering 
simulations of \cite{Barchas-2021-MNRAS,Hu-2022-ApJ}.  The discussion therein 
highlights the broader applicability of the results to neutron star settings.

\newpage

\section{Magnetic Thomson Transport using a Phase Matrix Approach}
 \label{sec:formalism}
The determination of the configuration of radiation anisotropy and polarization in magnetized 
plasmas can be addressed by standard radiative transfer techniques such as in 
\cite{Chandrasekhar-1960-book}.  It can also be approached using Monte Carlo simulation 
techniques, as it was in \cite{Barchas-2021-MNRAS} and \cite{Hu-2022-ApJ}.  Via either protocol, 
deep in the radiative transport environment, the radiation field realizes an asymptotic equilibrium 
wherein single scatterings do not change the distributions of photon angles and polarizations.  
This is the domain that is addressed here in deriving both analytic and numerical descriptions of 
the asymptotic radiation configuration.   Throughout the paper, the magnetic field \teq{\Bvec} 
threading the radiative transfer region is presumed uniform on scales much larger than the 
scattering mean free path.  

To determine the asymptotic state, it is necessary to include all information about polarizations and 
their interchange in scattering events.  To effect this, in this paper we employ the phase matrix 
approach for the 4 familiar Stokes parameters, \teq{I,Q,U,V}.  Here, \teq{I} represents the intensity, 
\teq{Q,U} define the linear polarization degree and angle configuration, and \teq{V} pertains to circular polarization.  
The equilibrium configuration is established via a redistribution function \teq{{\cal P}/\sigma} 
(to be defined) that specifies the probability of redistribution in angles and polarization states
under the operation of magnetic Thomson scattering.  The analysis is constructed using the 
\teq{4\times 4} phase matrix approach of \cite{Chou-1986-ApSS} that describes the mappings in Stokes parameter 
space in scatterings.  Such mappings correspond to dipole radiation by electrons gyrating 
in the magnetic field in response to the Newton-Lorentz force provided by the incoming electromagnetic
wave.  This classical description of magnetic Thomson scattering was first presented by \cite{Canuto-1971-PhRvD}.
Exploiting the azimuthal symmetry of the uniform magnetic field 
problem, without loss of generality we will eventually set \teq{U=0} and average over azimuthal angles
(phases) about \teq{\Bvec}.

\subsection{Radiative Transfer Set-up}

The photon wavenumber \teq{\kvec} (momentum/$\hbar$) will be described 
via polar \teq{(k,\,\theta, \,\phi)} coordinates in between scatterings, with 
\teq{\theta = \arccos (\kvechat\cdot\Bvechat )} being the polar angle relative to 
the field direction \teq{\Bvechat}, and \teq{\phi} being the azimuthal angle 
around the field.  In an unpolarized system, only the intensity \teq{I} comes into consideration,
and the radiative transfer equation assumes the form
\begin{equation}
  \dover{1}{c}\, \dover{dI}{dt} \; =\; - n_e \sigma (\kvec ) \, I(\kvec ) 
     + n_e \int  \dover{d\sigma (\kvec_i\to \kvec )}{d\Omega}  \, I(\kvec_i ) \, d\Omega_i \quad .
 \label{eq:intensity_RTE}
\end{equation}
Here, \teq{n_e} is the electron number density, and the differential cross section 
\teq{d\sigma/d\Omega} and accompanying total cross section are detailed in 
\cite{Canuto-1971-PhRvD}; see also \cite{Herold-1979-PhRvD} for the mathematically 
equivalent quantum version of the magnetic Thomson cross section.  The first term on the right 
of Eq.~(\ref{eq:intensity_RTE}) represents the rate of loss of photons from a given 
wavenumber bin \teq{\kvec}.  The second term on the right defines the rate of gain 
in a wavenumber bin, and integrates over all the directions \teq{d\Omega_i = d(\cos\theta_i)\, d\phi_i}
of the initial (pre-scattering) photons \teq{\kvec_i}; the subscript for the final photon is dropped.
For equilibrium in high opacity domains, the intensity is invariant, and so the left hand side is zero.  
Note that as the opacity increases, the approach to equilibrium is not uniform in either
direction or polarization due to the pathology of the cross section.
This is the focal regime of this paper, leading to the equality
\begin{equation}
   \sigma (\kvec ) \, I(\kvec ) \; =\; \int  \dover{d\sigma (\kvec_i\to \kvec )}{d\Omega} 
       \, I(\kvec_i ) \, d\Omega_i \quad .
 \label{eq:Stokes_equil_1D}
\end{equation}
The specific mathematical form of the magnetic Thomson cross section then 
determines the angular dependence of the intensity up to an unconstrained normalization.
This steady-state identity is appropriate for intensity evolution in a highly optically-thick 
plasma, and can routinely be adapted to introduce the photon polarization dependence.

Let \teq{\boldsymbol{S} = (I,Q,U,V) \equiv (S_1, S_2, S_3, S_4)} be the true Stokes vector
of the photon field that can be represented using the Poincar\'e sphere.  Using 
the complex vector representation of photon electric fields (polarizations) in \cite{Barchas-2021-MNRAS},
this Stokes vector can be written as
\begin{equation}
   \boldsymbol{S} \; \equiv\; 
   \left( \begin{array}{c}
   I \\[2pt]
   Q \\[2pt]
   U \\[2pt]
   V
   \end{array} \right) 
   \; =\; 
   \left( \begin{array}{c}
   \langle \calEthet \calEthet^* \rangle + \langle \calEphi \calEphi^* \rangle \\[2pt]
   \langle \calEthet \calEthet^* \rangle - \langle \calEphi \calEphi^* \rangle \\[2pt]
   \langle \calEthet \calEphi^* \rangle + \langle \calEthet^* \calEphi \rangle \\[2pt]
   i \, \langle  \calEthet \calEphi^* -  \calEthet^* \calEphi \rangle
   \end{array} \right) 
   \quad \hbox{with} \quad 
   \calEvec \; =\; {\cal E}_\theta\hat{\theta} + {\cal E}_\phi\hat{\phi}
   \;\equiv\; \bigl\vert \calEvec \bigr\vert \bigl( \hat{\cal E}_\theta\hat{\theta} + \hat{\cal E}_\phi\hat{\phi}  \bigr) 
 \label{eq:Stokes_polar_def}
\end{equation}
as the polarization vector in our polar coordinate specification.  The brackets 
\teq{\langle \dots \rangle} signify time averages of the products of wave field components;
the Stokes parameters can be summed over any number of photons.
One can also form a reduced Stokes parameter 3-vector
\teq{\hat{\boldsymbol{S}} = (\hat{Q}, \, \hat{U}, \, \hat{V}) \equiv (Q/I, \, U/I, \, V/I)} 
using the intensity to scale the other polarization quantities of interest.   
It then follows that \teq{\vert {\hat Q}\vert \leq 1} and similarly for \teq{{\hat U},{\hat V}}. 

Then extension of the radiative transfer equilibrium equation in Eq.~(\ref{eq:Stokes_equil_1D}) 
to treat polarized systems leads to the high opacity asymptotic identity
\begin{equation}
   \sigma (\kvec ) \, \boldsymbol{S}(\kvec ) 
   \; =\; \sum_{i,f=1,4} \int  \dover{d\sigma_{i\to f} (\kvec_i\to \kvec )}{d\Omega} 
       \, \boldsymbol{S}(\kvec_i ) \, d\Omega_i \quad .
 \label{eq:Stokes_equil_full}
\end{equation}
Here \teq{i\to f} represents the polarization transition in scattering from one initial 
photon polarization state \teq{i} to a final one \teq{f}.  Thus \teq{i,f=1\to 4} map 
through the Stokes parameters \teq{I,Q,U,V} in sequence.  This is a vector 
equation and the differential cross section is now a \teq{4\times 4} matrix, i.e. a tensor.
The transfer equation can also be cast in a form that uses the \teq{4\times 4} polarization 
phase matrix \teq{\mathbf{P}} with 16 elements \teq{P_{if}}:
\begin{equation}
   \sigma (\kvec ) \, \boldsymbol{S}(\kvec ) 
   \; =\; \int \mathbf{P} (\kvec_i\to \kvec ) \, \Bigl\lbrack  \sigma (\kvec_i )
       \, \boldsymbol{S}(\kvec_i )\Bigr\rbrack  \, d\Omega_i
   \quad ,\quad
   P_{if} \; =\;  \dover{1}{\sigma (\kvec_i)} \, 
      \dover{d\sigma_{i\to f} (\kvec_i\to \kvec )}{d\Omega} 
      \;\equiv\; \dover{r_0^2\, \hat{P}_{if}}{\sigma (\kvec_i)} \quad ,
 \label{eq:Stokes_equil_matrix}
\end{equation}
Accordingly, \teq{ \sigma (\kvec ) \, \boldsymbol{S}(\kvec ) } is an eigenvector of the 
phase matrix.  The scaled versions \teq{\hat{P}_{if}} of the phase matrix elements 
are introduced here to isolate their common cross section normalization.
Note that \teq{r_0=e^2/(m_ec^2)} is the classical electron radius, from which 
one derives the total non-magnetic Thomson cross section \teq{\sigt = 8\pi r_0^2/3}.

The Stokes vector \teq{\boldsymbol{S}(\kvec )} depends on the direction 
of the radiation relative to \teq{\Bvec}, thereby expressing its anisotropy.  The cross section 
\teq{\sigma (\kvec )} now depends on the polarization configuration, and therefore implicitly 
on \teq{\boldsymbol{S}(\kvec )}.  Eq.~(13) of \cite{Barchas-2021-MNRAS} presents the form
\begin{equation}
   \sigma (\kvec ) \; = \; \sigt \biggl\{ \SigmaB (\omega ) \,\hat{I} (\kvec )
       + \dover{1}{2} \Bigl[ 1 - \SigmaB (\omega ) \Bigr]  \Bigl\{ \hat{I} (\kvec ) 
       + \hat{Q} (\kvec ) \Bigr\} \, \bigl( 1- \mu^2\bigr)  
       + \DeltaB (\omega )\, \hat{V} (\kvec ) \, \mu \biggr\}
 \label{eq:sigma_tot_Stokes}
\end{equation}
for the polarized magnetic Thomson cross section, where \teq{\mu=\cos\theta} 
is the angle cosine of the initial photon, and \teq{\sigt} is the Thomson cross section.  
As above, the hat notation signifies \teq{\hat{Q} = Q/I}, etc, so that \teq{\hat{I}=1}.  Two 
functions appear that depend only on the photon frequency \teq{\omega = c/\vert \kvec\vert}:
\begin{equation}
    \SigmaB (\omega ) = \dover{\omega^2(\omega^2+\wcyc^2)}{ (\omega^2 - \wcyc^2)^2}
    \; =\; \dover{1}{2} \biggl\{ \dover{\omega^2}{(\omega - \wcyc)^2}  
         + \dover{\omega^2}{(\omega + \wcyc)^2} \biggr\} \; ,\quad 
   \DeltaB (\omega ) = \dover{2\omega^3\wcyc}{(\omega^2 - \wcyc^2)^2} 
   \;=\; \dover{1}{2} \biggl\{ \dover{\omega^2}{(\omega - \wcyc)^2} 
          - \dover{\omega^2}{(\omega + \wcyc)^2} \biggr\}
 \label{eq:SigmaB_DeltaB_def}
\end{equation}
that can be termed the {\sl linearity and circularity functions}, respectively.  They
depend only on the ratio of the photon frequency \teq{\omega} to 
the electron cyclotron frequency \teq{\wcyc = eB/(m_ec)}, and therefore 
are not independent functions.  The combination 
of angle and polarization dependence within \teq{\sigma (\kvec ) }
guarantees an intricate and frequency-dependent interplay between polarization and anisotropy in 
the transport of photons in a magnetized plasma.
Eq.~(\ref{eq:sigma_tot_Stokes}) agrees with Eq.~(4) of \cite{Whitney-1991-ApJS} 
and Eq.~(2.26) of \cite{Barchas-2017-thesis}, both of which 
were derived from the polarization phase matrix analysis of \cite{Chou-1986-ApSS}.
Note that in the limit of small fields, i.e. \teq{\omega\gg \wcyc}, then \teq{\SigmaB\to 1} and 
\teq{\DeltaB\to 0}, and Eq.~(\ref{eq:sigma_tot_Stokes}) reduces to the familiar 
non-magnetic Thomson result: the cross section is then independent of the 
magnetic field and the photon direction.

\subsection{Reduction of the System of Transport Equations}

The algebraic expressions that \cite{Chou-1986-ApSS} developed for the 
differential cross section are long and complicated in terms of their dependence 
on the polar angles \teq{\theta_{i,f}} and azimuthal/phase ones \teq{\phi_{i,f}}.
Moreover, they are then specialized to the case of a phase angle difference 
\teq{\phi_{fi}= \phi_f - \phi_i=0}, which does not capture all the required information for 
the radiative transfer analysis.  Accordingly, we use as our starting point the modified 
version posited in Eq.~(2) of \cite{Whitney-1991-ApJS}; see also Eq.~(2.24) and 
subsequent equations on pages 42-43 of \cite{Barchas-2017-thesis}.
While \cite{Chou-1986-ApSS} formed a representation of the phase matrix using the true Stokes 
vectors \teq{\boldsymbol{S}},  \cite{Whitney-1991-ApJS} noted that it proved 
algebraically more convenient to employ the decomposition
\teq{I= I_{\parallel}+I_{\perp}} and \teq{Q= I_{\parallel}-I_{\perp}} 
and work with \teq{I_{\parallel,\perp}}.  This amounts to a simple rotation in the 
\teq{(S_1, S_2)} elements that can be applied to the phase matrix/tensor.  
By inspection of the Stokes parameter definitions in Eq.~(\ref{eq:Stokes_polar_def}), 
one quickly discerns that \teq{I_{\parallel} = \langle \calEthet \calEthet^* \rangle} 
and so contains information about electric field components in the \teq{\kvec - \Bvec} plane.
Thus, \teq{I_{\parallel}} represents the intensity of the \teq{\parallel} linear polarization 
component, the so-called ordinary mode.  Similarly, \teq{I_{\perp} = \langle \calEphi \calEphi^* \rangle} 
only contains information on electric field components perpendicular to the \teq{\kvec - \Bvec} plane, 
and thus defines the intensity of the \teq{\perp} linear polarization (extraordinary mode).
After an appropriate amount of algebraic development,
eventually we will revert to the Chou representation 
for aesthetic reasons that will shortly become apparent.

Accordingly, we seek to algebraically develop the equilibrium integral equation problem
\begin{equation}
   \boldsymbol{\Lambda}  (\kvec ) 
   \; =\; \int \mathbf{P} (\kvec_i\to \kvec ) \, \boldsymbol{\Lambda}  (\kvec_i ) \, d\Omega_i
  \quad ,\quad 
  \boldsymbol{\Lambda} (\kvec ) \;\equiv\; \sigma (\kvec ) \, \boldsymbol{S}(\kvec ) \quad ,
 \label{eq:Stokes_equil_penult}
\end{equation}
with \teq{\boldsymbol{S}\to (I_{\parallel},\, I_{\perp}, \, U, \, V)}, and reduce it in 
preparation for numerical solution.  The first step is to adopt a random phase approximation.
At high opacity, uniformity in azimuthal ``phase'' \teq{\phi} about \teq{\Bvec} should be 
realized because the differential cross section has no peculiar dependence on 
the azimuthal angles; i.e. the configuration is invariant under rotations about \teq{\Bvec}.  
This then applies to both \teq{\boldsymbol{S}} 
and also \teq{\boldsymbol{\Lambda }}, since \teq{\sigma (\kvec )} has no 
explicit dependence on azimuth.  Thus, \teq{\boldsymbol{\Lambda}  (\kvec )
\to \boldsymbol{\Lambda}  (\omega ,\, \mu )}, for photon frequency 
\teq{\omega} and propagation angle cosine \teq{\mu = \cos\theta}
relative to \teq{\Bvec}, and we shall employ the \teq{\sigma (\kvec ) \to \sigma  (\omega ,\, \mu )}   
correspondence in what follows.  Also, without loss of generality (WLOG), one can choose 
polar coordinates so that the Stokes U is zero and Stokes Q provides the only 
contribution to the linear polarization.  This is tantamount to an orientation of the 
polar axis along \teq{\Bvec} and a particular choice for the zero of the 
azimuthal angle coinciding with the \teq{\kvec - \Bvec} plane.  This sets 7 elements of the phase 
matrix to be zero.  With this simplification, the system can be expressed using 
\teq{3\times 3} reduced phase matrix \teq{\mathbf{P}_{\rm red}}.  The transport equilibrium is 
then expressed via
\begin{equation}
   \boldsymbol{\Lambda}_{\parallel, \perp}   (\omega ,\, \mu ) 
   \; =\; \int_{-1}^1 {\cal R} (\mu_i, \mu ) \, \boldsymbol{\Lambda}_{\parallel, \perp}    (\omega ,\, \mu_i ) \, d\mu_i 
   \quad \hbox{with}\quad
   \boldsymbol{\Lambda}_{\parallel, \perp}   (\omega ,\, \mu ) \; =\; \sigma  (\omega ,\, \mu )
   \left( \begin{array}{c}
       I_{\parallel}\\
       I_{\perp}\\
       V\\
   \end{array} \right) \;\equiv\;
   \left( \begin{array}{c}
       \Lambda_{\parallel} (\omega ,\, \mu )\\
       \Lambda_{\perp} (\omega ,\, \mu )\\
       \Lambda_V (\omega ,\, \mu )\\
   \end{array} \right) \;\; ,
 \label{eq:Lambda_mat_form}
\end{equation}
constituting a three-dimensional eigenvalue/eigenvector problem.  Observe that in 
Eq.~(\ref{eq:Lambda_mat_form}), the angle cosine \teq{\mu} represents that for the 
final (scattered) photon on the right hand side, and the initial pre-scattering photon on 
the left side.  The subscript \teq{\parallel, \perp} marks that this applies to the 
\cite{Whitney-1991-ApJS} mix of polarizations.  Here we have now introduced 
the {\bf re-distribution phase matrix} mapping, defined for a \teq{\kvec_i\to \kvec_f}
scattering via
\begin{equation}
   {\cal R} (\mu_i,  \mu_f ) \; =\; \int_0^{2\pi} \mathbf{P}_{\rm red} (\kvec_i\to \kvec_f ) \, d\phi_{fi}
   \; \equiv\; \int_0^{2\pi} 
   \left( \begin{array}{ccc}
      P_{11} & P_{12} & P_{14}\\
      P_{21} & P_{22} & P_{24}\\
      P_{41} & P_{42} & P_{44}\\
   \end{array} \right) \, d\phi_{fi} \; ,
 \label{eq:redistrib_def}
\end{equation}
a \teq{3\times 3} matrix representation of the \teq{U=0} specialization of the full phase space matrix;
\teq{U\neq 0} circumstances are easily retrievable via a rotation of coordinates.
Herein, we will label the resulting elements via their Stokes parameter 
identification, so that \teq{1\to I_{\parallel}}, \teq{2\to I_{\perp}}, \teq{4\to V}.  This reduced
\teq{3\times 3} matrix (subscripted ``red'') will be employed in seeking an 
eigenvalue/eigenvector solution to the radiative transfer equation.  
Note that the azimuthal dependence of the full phase matrix is captured in simple trigonometric functions 
of \teq{\phi_{fi}=\phi_f-\phi_i}, so this serves as a suitable integration variable for the phases.

Using the full expressions for the scaled phase matrix elements \teq{\hat{P}_{if}} given 
in Appendix A (see Eq.~(\ref{eq:Stokes_equil_matrix}) for their definition), the  integration of 
the \teq{P_{if}} over \teq{\phi_{fi}} leads to the explicit form for \teq{{\cal R} (\mu_i, \, \mu_f)}:
\begin{equation}
   {\cal R} (\mu_i, \, \mu_f)\; =\;  \dover{\pi r_0^2}{\sigma  (\omega ,\, \mu_i )} 
   \left( \begin{array}{ccc}
       2 \bigl( 1 - \mu_i^2 \bigr) \bigl( 1 - \mu_f^2 \bigr) 
        + \SigmaB  \, \mu_i^2\, \mu_f^2 
             & \SigmaB   \, \mu_f^2 & \DeltaB   \, \mu_i\, \mu_f^2 \\[4pt]
      \SigmaB   \, \mu_i^2 & \SigmaB   &  \DeltaB   \, \mu_i\\[2pt]
      2\DeltaB   \, \mu_i^2 \, \mu_f & 2\DeltaB   \, \mu_f & 2\SigmaB   \, \mu_i \, \mu_f \\
   \end{array}  \right) \; ,
 \label{eq:calR_matrix}
\end{equation}
where the functional dependence of the \teq{\SigmaB (\omega )} and \teq{\DeltaB (\omega )} 
is implied, here and hereafter.  Following \cite{Whitney-1991-ApJS}, one can now form the 
partial cross sections for all the polarization elements by summing the first two rows of each 
column in \teq{\sigma  (\omega ,\, \mu_i ) \, {\cal R}}, and integrating \teq{\mu_f} over the 
range \teq{-1 \leq \mu_f \leq 1}.  These two rows are required since they sum to yield the 
total intensity.  The three columns respectively produce the contributions to the produced 
\teq{I_{\parallel}}, \teq{I_{\perp}} and \teq{V} in this mixed polarization configuration.  Thus,
\begin{equation}
   \sigma_{\parallel}  (\omega ,\, \mu_i ) \; = \; \sigt \Bigl( 1 - \mu_i^2 + \SigmaB  \, \mu_i^2\Bigr)
   \quad ,\quad
   \sigma_{\perp}  (\omega ,\, \mu_i ) \; = \; \sigt \SigmaB
   \quad ,\quad
   \sigma_V  (\omega ,\, \mu_i ) \; = \; \sigt \DeltaB \mu_i \quad .   
 \label{eq:sig_pol_comps}
\end{equation}
The first two reproduce the corresponding forms in Eq.~(6) of \cite{Whitney-1991-ApJS},
wherein we note that there is a factor of 4 typographical error (too large) in her result 
for \teq{\sigma_V}.  The total polarized cross section that appears in Eq.~(\ref{eq:calR_matrix})
is then
\begin{equation}
   \sigma  (\omega ,\, \mu_i ) \;\equiv\;  \sigma (\kvec )
   \; =\;  \sigma_{\parallel} \hat{I}_{\parallel} + \sigma_{\perp}  \hat{I}_{\perp} + \sigma_V \hat{V}
   \; =\; \sigt \biggl\{ \Bigl( 1 - \mu_i^2 + \SigmaB  \, \mu_i^2\Bigr) \hat{I}_{\parallel}
      + \SigmaB \hat{I}_{\perp} + \DeltaB \mu_i\, \hat{V} \biggr\} \quad ,
 \label{eq:sigma_tot_Stokes_mix}
\end{equation}
which is equivalent to Eq.~(\ref{eq:sigma_tot_Stokes}).  The specific mathematical form 
for \teq{\sigma  (\omega ,\, \mu_i ) } that results from the equilibrium solution is 
ultimately given in Eq.~(\ref{eq:sigma_tot_asymptotic}) below.

\subsection{Symmetrized Radiative Transfer Equations} 
 \label{sec:symmetrzied_FTE}

The re-distribution phase matrix in Eq.~(\ref{eq:calR_matrix}) is not symmetric.  
Such character results from the mixing of linear polarizations \teq{I_{\parallel} = (I+Q)/2}
and \teq{I_{\perp} = (I-Q)/2}.   While the algebra is more compact for this configuration, 
the final true polarization information can be recovered at the end, 
via \teq{I = I_{\parallel} + I_{\perp}} and \teq{Q = I_{\parallel} - I_{\perp}}.  This mix obscures a fundamental 
symmetry, which is only apparent in the pure Stokes formulation.  Therefore, to 
elucidate it, we invert the mixing, which in matrix form satisfies
\begin{equation}
    \left( \begin{array}{c}
       I_{\parallel}\\
       I_{\perp}\\
       U\\
       V\\
    \end{array} \right) \; = \; \boldsymbol{\rho} \cdot
    \left( \begin{array}{c}
       I\\
       Q\\
       U\\
       V\\
    \end{array} \right) 
    \quad \hbox{for}\quad
    \boldsymbol{\rho} \; =\;  \left( \begin{array}{cccc}
       1/2 & 1/2 & \; 0 \; & \; 0\\
       1/2 & -1/2 & \; 0 \; & \; 0\\
       0 & 0 & \; 1 \; & \; 0\\
       0 & 0 & \; 0 \; & \; 1\\
    \end{array} \right) \quad .
 \label{eq:Stokes_mix}
\end{equation}
Here, \teq{\boldsymbol{\rho}} is the pertinent Stokes parameter ``mixing matrix.''
This matrix can be used to transcribe the full scattering phase matrix \teq{ \mathbf{P}} 
via \teq{ {\cal M} \;\propto\; \boldsymbol{\rho}^{-1} \cdot \mathbf{P} \cdot \boldsymbol{\rho}}, 
so that the true Stokes phase matrix is now used to generate a form \teq{{\cal M} (\mu_i, \, \mu_f)} 
for  \teq{{\cal R} (\mu_i, \, \mu_f)} to be used in an analog of Eq.~(\ref{eq:Lambda_mat_form}).  
This form is defined to be 
\begin{equation}
  {\cal M} (\mu_i,  \mu_f ) \; =\;  \dover{\sigma  (\omega ,\, \mu_i )}{\pi r_0^2} 
   \int_0^{2\pi} \Bigl[ \boldsymbol{\rho}^{-1} \cdot \mathbf{P} \cdot \boldsymbol{\rho} \Bigr]_{\rm red} \, d\phi_{fi}
   \; =\; 
   \left( \begin{array}{ccc}
      {\cal M}_{11} & {\cal M}_{12} & {\cal M}_{14}\\
      {\cal M}_{21} & {\cal M}_{22} & {\cal M}_{24}\\
      {\cal M}_{41} & {\cal M}_{42} & {\cal M}_{44}\\
   \end{array} \right) \quad .
 \label{eq:calM_mat_def}
\end{equation}
Note that here, as with Eq.~(\ref{eq:redistrib_def}), the subscript ``red'' notation implies 
a reduced \teq{3\times 3} matrix format that eliminates the trivial \teq{U=0} contribution to 
the radiation transport.  After integration over azimuths, all the \teq{{\cal M}_{i3}} and 
\teq{{\cal M}_{3j}} are identically zero.  Defining \teq{\kappa = (1-\mu_i^2)(1 - \mu_f^2)},
which captures the non-magnetic portion of the differential cross section,  
the remaining non-zero elements are as follows:
\begin{equation}
   \left( \begin{array}{ccc}
      {\cal M}_{11} & {\cal M}_{12} & {\cal M}_{14}\\
      {\cal M}_{21} & {\cal M}_{22} & {\cal M}_{24}\\
      {\cal M}_{41} & {\cal M}_{42} & {\cal M}_{44}\\
   \end{array} \right) 
   \; =\; 
   \left( \begin{array}{ccc}
       \kappa + \dover{\SigmaB}{2}  \, \bigl( 1 + \mu_i^2 \bigr) \bigl( 1 + \mu_f^2 \bigr) 
       & \kappa -  \dover{\SigmaB}{2} \bigl( 1-\mu_i^2 \bigr) \bigl( 1 + \mu_f^2 \bigr)
       &  \DeltaB \mu_i \bigl( 1 + \mu_f^2 \bigr) \\[6pt]
       \kappa -  \dover{\SigmaB}{2} \bigl( 1-\mu_f^2 \bigr) \bigl( 1 + \mu_i^2 \bigr)
       & \kappa \biggl[ 1 + \dover{\SigmaB}{2} \biggr]
       &   - \DeltaB \mu_i \bigl( 1 - \mu_f^2 \bigr) \\[6pt]
        \DeltaB \mu_f \bigl( 1 + \mu_i^2 \bigr)
       & - \DeltaB \mu_f \bigl( 1 - \mu_i^2 \bigr)
       & 2\SigmaB   \, \mu_i \, \mu_f \\
   \end{array}  \right) \; .
 \label{eq:redistrib_calM_def}
\end{equation}
Observe that the cross section dependence has been extracted outside 
the definition of \teq{{\cal M}} so as to simplify its form.  These matrix elements are 
identified in Eq.~(11) of \cite{Caiazzo-2021-MNRAS}, modulo the sign convention pertaining 
to the \teq{\DeltaB} elements that is discussed at the end of Appendix A.

Inspection of these elements reveals an important symmetry: transposition of the matrix 
in conjunction with the interchange \teq{\mu_i \leftrightarrow\mu_f} generates the 
original form of \teq{\cal M}.  This captures {\sl the time-reversal symmetry of the scattering 
process in the Thomson limit}, a property that is not explicitly apparent for Whitney's 
choice of the polarization decomposition.  Accordingly, the pure Stokes parameter 
decomposition delivered in this Subsection is preferred, on aesthetic grounds.
Yet, the end product solutions obtained in Sec.~\ref{sec:solutions} below do not 
depend on this choice.

Another important feature is that \teq{{\cal M}} is a \teq{3\times 3} matrix of determinant zero
(so too is \teq{{\cal R}}).  
This is essentially a consequence of it constituting a scaling of the imaginary part of the 
dispersion tensor \teq{\boldsymbol{\Delta}} that is derived from the dielectric response tensor 
for a magnetized plasma in the zero temperature, non-relativistic limit \citep[see][]{Ichimaru-1973-book}.  
The dispersion tensor has zero determinant because its real part describes normal modes of 
electromagnetic wave propagation in the plasma medium in the absence of driver 
currents, with its imaginary part capturing the absorptive (scattering) contribution.  
For transverse electromagnetic waves, \teq{\vert\boldsymbol{\Delta}\vert =0} 
guarantees that there are two normal modes, and also that there are only two 
linearly independent polarization contributions to the radiative transfer incurred 
by Thomson scattering, as will soon become evident.  

As before, the forms for the scattering matrix elements in Eq.~(\ref{eq:redistrib_calM_def})
may be further integrated over \teq{\mu_f} to obtain the various polarization contributions 
to the total cross section.  Since this polarization configuration is unmixed, only the 
first row is relevant, since the end product is just the total intensity.  Thus,
\begin{eqnarray}
   \sigma_{\hbox{\sixrm I}} & = & \pi r_0^2 \int_{-1}^1 {\cal M}_{11} \, d\mu_f
   \; =\; \dover{\sigt}{2} \Bigl\{ \bigl( 1 - \mu_i^2 \bigr) 
        + \SigmaB  \bigl( 1 + \mu_i^2 \bigr) \Bigr\} \nonumber\\[2.0pt]
   \sigma_{\hbox{\sixrm Q}} & = & \pi r_0^2 \int_{-1}^1 {\cal M}_{12} \, d\mu_f
   \; =\; \dover{\sigt}{2} \,  \bigl( 1 - \mu_i^2 \bigr) \, \bigl[ 1 - \SigmaB \bigr]
 \label{eq:sigpol_unmixed} \\[2.0pt]
   \sigma_{\hbox{\sixrm V}} & = & \pi r_0^2 \int_{-1}^1 {\cal M}_{14} \, d\mu_f
   \; =\; \sigt \,\DeltaB\, \mu_i\quad ,\nonumber
\end{eqnarray}
with \teq{\sigma_{\hbox{\sixrm U}}=0} trivially.  The total polarized cross section is then obtained via 
\teq{ \sigma \; =\;  \sigma_{\hbox{\sixrm I}} \hat{I} + \sigma_{\hbox{\sixrm Q}}  \hat{Q} + \sigma_{\hbox{\sixrm V}} \hat{V}}, 
and yields Eq.~(\ref{eq:sigma_tot_Stokes}) exactly.

With this time-symmetric construction, the updated system of equations to describe the equilibrium polarization configuration 
\teq{\boldsymbol{\Lambda}  (\omega ,\, \mu ) \equiv \sigma  (\omega ,\, \mu )\times (I,\, Q, \, V)} 
takes the form
\begin{equation}
   \left( \begin{array}{c}
      \Lambda_I(\omega ,\, \mu )\\[2pt]
      \Lambda_Q(\omega ,\, \mu )\\[2pt]
      \Lambda_V(\omega ,\, \mu )\\
   \end{array} \right)
   \; =\; \pi r_0^2 \int_{-1}^1 
   \left( \begin{array}{ccc}
      {\cal M}_{11} & {\cal M}_{12} & {\cal M}_{14}\\
      {\cal M}_{21} & {\cal M}_{22} & {\cal M}_{24}\\
      {\cal M}_{41} & {\cal M}_{42} & {\cal M}_{44}\\
   \end{array} \right) 
   \left( \begin{array}{c}
      \Lambda_I(\omega ,\, \mu_i )\\[2pt]
      \Lambda_Q(\omega ,\, \mu_i )\\[2pt]
      \Lambda_V(\omega ,\, \mu_i )\\
   \end{array} \right) \, \dover{d\mu_i}{{\sigma  (\omega ,\, \mu_i )}}  \quad ,
 \label{eq:Lambda_mat_Stokes}
\end{equation}
with the \teq{{\cal M}_{ij}} matrix elements given by Eq.~(\ref{eq:redistrib_calM_def}).
This is the basic staging platform for further reduction to hone the system so 
that it is ready for generating numerical and analytic solutions.  Hereafter, as in 
Eq.~(\ref{eq:Lambda_mat_form}), the usage of \teq{\mu_i} on the left and \teq{\mu_f} 
inside the integrals has been deprecated since the angle cosine \teq{\mu} represents 
that for the final (scattered) photon on the right hand side, and the initial pre-scattering 
photon on the left.

\subsection{A Distilled Dual System of Linearly-Independent Scattering Transfer Equations} 

One final stage of analytic reduction is required before proceeding to 
identifying the solutions.  As a result of the simple quadratic dependence of 
the \teq{{\cal M}_{ij}} in Eq.~(\ref{eq:redistrib_calM_def}), an
inspection of Eq.~(\ref{eq:Lambda_mat_Stokes}) quickly reveals that the 
\teq{\Lambda_i} functions must possess the following simple quadratic 
forms:
\begin{equation}
   \Lambda_I(\mu ) \; =\; {\cal N} + {\cal A} \mu^2
   \quad ,\quad 
   \Lambda_Q(\mu ) \; =\; {\cal L} \bigl( \mu^2 - 1 \bigr)
   \quad ,\quad 
   \Lambda_V(\mu ) \; =\; 2 {\cal C} \mu \quad .
 \label{eq:calNALC_def}
\end{equation}
The coefficients \teq{\cal N} (normalization),  \teq{\cal A} (anisotropy), 
\teq{\cal L} (polarization linearity) and  \teq{\cal C} (polarization circularity) 
depend only on \teq{\omega /\wcyc}.  
We therefore proceed to solve the system of Fredholm equations of the second kind
as a Neumann series problem that is truncated at two 
terms due to the quadratic character of the solutions, and is therefore 
an exact, closed problem.  We write it as
\begin{equation}
   \left( \begin{array}{c}
      {\cal N} + {\cal A} \, \mu^2\\[2pt]
      {\cal L} \bigl( \mu^2 - 1 \bigr) \\[2pt]
      2 {\cal C} \, \mu \\
   \end{array} \right)
   \; =\; \pi r_0^2 \int_{-1}^1   
   \left( \begin{array}{ccc}
      {\cal M}_{11} & {\cal M}_{12} & {\cal M}_{14}\\
      {\cal M}_{21} & {\cal M}_{22} & {\cal M}_{24}\\
      {\cal M}_{41} & {\cal M}_{42} & {\cal M}_{44}\\
   \end{array} \right) 
   \left( \begin{array}{c}
      {\cal N} + {\cal A} \, \mu_i^2\\[2pt]
      {\cal L} \bigl( \mu_i^2 - 1 \bigr)\\[2pt]
      2 {\cal C} \, \mu_i \\
   \end{array} \right) \, \dover{d\mu_i}{{\sigma  (\omega ,\, \mu_i )}} \quad ,
 \label{eq:redistribute_Neumann}
\end{equation}
with the objective of solving for the coefficients \teq{\cal N,A,L,C}.  Two 
simplifications appear promptly.  First, the normalization of the solution 
space is arbitrary, so all the coefficients will scale as the value of 
\teq{\cal N}.  So WLOG, we set \teq{{\cal N}=1}.  Next, by inspection of 
Eq.~(\ref{eq:redistribute_Neumann}), one quickly discerns that the 
coefficient of \teq{\mu^2} in the first row of the RHS of Eq.~(\ref{eq:redistribute_Neumann})
is identical to that of \teq{\mu^2-1} in the second row of Eq.~(\ref{eq:redistribute_Neumann}).
This yields the identity \teq{{\cal L} = {\cal A}}, i.e. that the polarization linearity coincides 
with the intensity anisotropy.  Accordingly, this leaves only 
two undetermined coefficients, \teq{\cal A} and \teq{\cal C} in the system of integral equations.

The system can be re-written by exploring the \teq{\mu} dependence on both sides
of each row.  Evaluating the first row for 
\teq{\mu =1} gives
\begin{equation}
   1 + {\cal A} \; = \; \dover{3}{4} \int_0^1  \dover{\sigt\, d\mu_i}{\sigma  (\omega ,\, \mu_i )} 
   \biggl\{ \SigmaB \Bigl[  1 + {\cal A} +  \mu_i^2 \bigl(  1 - {\cal A} + 2{\cal A} \, \mu_i^2 \bigr) \Bigr]
         + 4 \DeltaB {\cal C} \mu_i^2   \biggr\} \;\; .
 \label{eq:solution1}
\end{equation}
Observe that we have used the even nature of the integrands to halve the integration interval.
For a second, seemingly independent equation, we evaluate the first two 
rows of Eq.~(\ref{eq:redistribute_Neumann}) for \teq{\mu =0} and add the results.  This yields
\begin{equation}
    1 - {\cal A} \; = \; \dover{3}{2} \int_0^1  \dover{\sigt\, d\mu_i}{\sigma  (\omega ,\, \mu_i )} 
       \bigl( 1 - \mu_i^2 \bigr)  \Bigl[ 1 - {\cal A} + 2 {\cal A} \, \mu_i^2 \Bigr] \quad .
 \label{eq:solution2}
\end{equation}
The  bottom row of Eq.~(\ref{eq:redistribute_Neumann}) yields an additional constraint, 
\begin{equation}
   2\, {\cal C} \; =\;  \dover{3}{4} \int_0^1  \dover{\sigt\, d\mu_i}{\sigma  (\omega ,\, \mu_i )} 
   \biggl\{  \DeltaB \Bigl[  1 + {\cal A} +  \mu_i^2 \bigl(  1 - {\cal A} + 2{\cal A} \, \mu_i^2 \bigr) \Bigr]
      + 4 \SigmaB {\cal C} \, \mu_i^2 \biggr\}  \; .
 \label{eq:solution3}
\end{equation}
Using the quadratic forms for the \teq{\Lambda_i}, since the intensity \teq{I} is proportional to
\teq{1 + {\cal A}\mu^2}, one can quickly deduce the scaled forms of the Stokes parameters:
\begin{equation}
   \hat{I} \; =\; 1
   \quad ,\quad
   \hat{Q} \; =\; \dover{{\cal A} \bigl( \mu^2 - 1\bigr)}{1 + {\cal A} \mu^2}
   \quad ,\quad
   \hat{V} \; =\; \dover{2{\cal C} \mu}{1 + {\cal A} \mu^2} \quad .
 \label{eq:hatIQV}
\end{equation}
These mathematical forms were numerically deduced in the Monte Carlo simulation analysis 
presented in \cite{Barchas-2021-MNRAS} pertaining to the high opacity radiation transport configuration 
appropriate deep inside ionized neutron star atmospheres.
The requirements that \teq{\vert \hat{Q}\vert , \vert \hat{V}\vert \leq 1} impose the physical constraints
\teq{-1 < {\cal A} \leq 1} and \teq{0 < {\cal C} < 1} on the parameters.  The partner Stokes 
parameters  in the \cite{Whitney-1991-ApJS} representation that are germane to linear polarization are then
\begin{equation}
   \hat{I}_\parallel \; =\; \dover{1}{2} \bigl( \hat{I} + \hat{Q} \bigr) 
   \; =\; \dover{1+ {\cal A} \bigl( 2\mu^2 - 1\bigr)}{2 \bigl( 1 + {\cal A} \mu^2\bigr)}
   \quad ,\quad
   \hat{I}_\perp \; =\; \dover{1}{2} \bigl( \hat{I} - \hat{Q} \bigr) 
   \; =\; \dover{1+ {\cal A}}{2\bigl(1 + {\cal A} \mu^2 \bigr)} \quad .
 \label{eq:hatI_Whitney}
\end{equation}
From this, one can conclude that in the non-magnetic domain of \teq{\omega\gg \wcyc}, 
the anisotropy should go to zero, \teq{{\cal A}\to 0}, and the system is unpolarized with
\teq{\hat{I}_\parallel = 1/2 = \hat{I}_\perp}.  In the opposite extreme, the highly-magnetic 
\teq{\omega\ll \wcyc} domain, we will eventually discern that \teq{{\cal A}\to -1} and 
thus \teq{\hat{I}_\parallel = 1} with \teq{\hat{I}_\perp = 0} for most (but not all) \teq{\mu}; 
the normal linear polarization dominates the radiation configuration, the circumstance 
for X rays within magnetar atmospheres.

Inserting Eq.~(\ref{eq:hatIQV}) into Eq.~(\ref{eq:sigma_tot_Stokes}), it follows that
the magnetic Thomson cross section in the high opacity domain then takes the form 
\begin{equation}
   \sigma (\omega ,\, \mu ) \; = \; \sigt \biggl\{ \SigmaB  + \dover{1 - \SigmaB}{2}  
       \dover{ 1- {\cal A} + 2 {\cal A} \mu^2}{ 1 + {\cal A}\mu^2} \, \bigl( 1- \mu^2\bigr) 
    + \DeltaB \, \dover{2{\cal C} \mu^2}{ 1 + {\cal A}\mu^2} \biggr\} \quad .
 \label{eq:sigma_tot_asymptotic}
\end{equation}
As always, the functional dependence of the \teq{\SigmaB (\omega )} and \teq{\DeltaB (\omega )} 
is implied.  Note that, as expected, \teq{ \sigma (\omega ,\, \mu )} is invariant under the reflection 
\teq{\mu \to -\mu}.  This expression for the total cross section can be used to replace the 
\teq{\DeltaB {\cal C}} term in the integrand of Eq.~(\ref{eq:solution1}).  After a modicum 
of algebra, this manipulation of Eq.~(\ref{eq:solution1}) reproduces Eq.~(\ref{eq:solution2}).  
Therefore these two equilibrium transfer equations are linearly dependent.
This redundancy is a direct consequence of the determinants of the \teq{{\cal M}}
and \teq{{\cal R}} matrices both being zero, thereby precisely capturing the information 
on the two transverse-mode eigenfunctions of the plasma dispersion tensor, as 
discussed in Sec.~\ref{sec:symmetrzied_FTE} above.

Therefore, Eqs.~(\ref{eq:solution1}) and~(\ref{eq:solution3}) constitute two 
linearly independent integral equations in the two parameters \teq{\cal A}
and \teq{\cal C}, and so this establishes a well-posed, closed system for solution.
We now write this dual system in the form
\begin{equation}
   1 + {\cal A} \; = \; \dover{3}{4} \int_0^1  \dover{\sigt\, d\mu_i}{\sigma  (\omega ,\, \mu_i )} 
    \Bigl\{  \SigmaB \, {\cal K}(\mu_i) + 4 \DeltaB {\cal C} \, \mu_i^2 \Bigr\} 
   \qquad ,\qquad   
   2 \, {\cal C} \; = \;  \dover{3}{4} \int_0^1  \dover{\sigt\, d\mu_i}{\sigma  (\omega ,\, \mu_i )} 
   \Bigl\{  \DeltaB \, {\cal K}(\mu_i)  + 4 \SigmaB {\cal C} \, \mu_i^2 \Bigr\} \quad ,
  \label{eq:master_system}
\end{equation}
where we have the definition and the identity 
\begin{equation}
   {\cal K}(\mu_i) \; \equiv \; 1 + {\cal A} +  \mu_i^2 \bigl(  1 - {\cal A} + 2{\cal A} \, \mu_i^2 \bigr)
   \; =\; \bigl( 1 + {\cal A}\bigr) \bigl(1 - \mu_i^2 \bigr)  +  2 \mu_i^2 \bigl(  1 + {\cal A} \, \mu_i^2 \bigr)\quad .
 \label{eq:integrand_SigDel_coeff}
\end{equation}
These equilibrium radiative transfer equations clearly display the circular 
polarization parity property that \teq{{\cal C} \to - {\cal C}} when \teq{\DeltaB \to - \DeltaB}.
Various linear combinations of the identities in Eq.~(\ref{eq:master_system}) can be formed 
to derive alternative dual systems of integral equations, yet without material improvement 
in facilitating their numerical solution.  Accordingly, Eq.~(\ref{eq:master_system}) defines 
our baseline doublet of master equations for the radiative transport in high magnetic Thomson opacity, 
whose solutions for \teq{{\cal A}(\omega/\wcyc )} and \teq{{\cal C}(\omega/\wcyc )} we will develop.
These solutions will encapsulate all the polarization and anisotropy character for the two
normal electromagnetic modes in highly opaque magnetized plasma.

\section{Transport Solutions in the High Opacity Domain}
 \label{sec:solutions}

The master equations in Eq.~(\ref{eq:master_system}) are transcendental in the
parameters \teq{{\cal A}(\omega/\wcyc )} and \teq{{\cal C}(\omega/\wcyc )} and 
so must be solved numerically in general.  Yet, in select frequency domains, 
analytic solution is possible, and desirable, and so these are pursued first before 
addressing the general numerical solutions and empirical fitting functions 
for the frequency dependence of the parameter solutions.

\subsection{Special Cases}
 \label{sec:special} 

There are four special frequencies/ranges of interest to the developments.  The first is the 
low frequency domain, \teq{\omega\ll \wcyc}, which is particularly germane to the 
surface X-ray emission of magnetars.  Since both \teq{\SigmaB\ll 1} and \teq{\DeltaB\ll 1}
when \teq{\omega/\wcyc\ll 1}, it is immediately apparent the two integrals on the 
right of Eq.~(\ref{eq:master_system}) are both small, requiring that 
\teq{{\cal A}\approx -1} and \teq{{\cal C}\approx 0}.  Given the forms in 
Eq.~(\ref{eq:SigmaB_DeltaB_def}), it can be deduced that both \teq{1+{\cal A}} and \teq{{\cal C}} 
are of order \teq{O(\omega/\wcyc )} or smaller. Yet we wish to be more precise 
in specifying their behaviors with frequency.  To this end, we first examine the 
cross section contribution to the denominators of the integrands:  in 
Eq.~(\ref{eq:sigma_tot_asymptotic}), both the \teq{\DeltaB {\cal C}} and \teq{\SigmaB}
terms can be neglected, and \teq{{\cal A}\to -1} can be set in remaining dominant terms.
In the numerator, the form of \teq{\cal K} can be reduced using \teq{{\cal A}\to -1}.  These two 
specializations yield
\begin{equation}
    \dover{\sigma  (\omega ,\, \mu_i )}{\sigt} \;\approx\; (1-\mu_i^2) + O\Bigl[ (\omega /\wcyc )^2\bigr]
    \quad ,\quad
    {\cal K}(\mu_i) \;\approx\; 2\mu_i^2 (1-\mu_i^2) + O\Bigl[ (\omega /\wcyc )^2\bigr] \quad .
 \label{eq:sigma_calK_lowfreq}
\end{equation}
The \teq{\cal C} terms in the numerator contribute negligibly at these low frequencies, 
resulting in a simple evaluation of the two integrals.  From this, one determines
\begin{equation}
   1 + {\cal A} \;\approx\; \dover{\SigmaB}{2} \; \approx \dover{1}{2} \left( \dover{\omega}{\wcyc} \right)^2
   \quad ,\quad
   {\cal C} \;\approx\; \dover{\DeltaB}{4} \; \approx\;  \dover{1}{2} \left( \dover{\omega}{\wcyc} \right)^3
   \quad \hbox{for} \quad  \dover{\omega}{\wcyc} \;\ll\; 1\quad .
 \label{eq:calA_calC_lowfreq}
\end{equation}
With a considerable amount of work, the next order corrections to these asymptotic behaviors 
can be determined.  Yet there is little need for retaining such orders, as 
Eq.~(\ref{eq:calA_calC_lowfreq}) suffices in helping validate the numerics 
in Sec.~\ref{sec:numerics}.

The \teq{\omega\ll \wcyc} domain is pertinent to X rays present in magnetar 
atmospheres.  The small value of \teq{\cal C} implies minimal circular polarization 
as the photon frequency is remote from the cyclotron one.  From Eq.~(\ref{eq:hatIQV}), 
as \teq{{\cal A}\approx -1}, for most values of \teq{\mu}, \teq{\hat{Q}\approx 1} so that 
the radiation configuration is strongly linearly polarized with a dominance of the 
\teq{\parallel} mode.  Yet the small departure of \teq{\cal A} from -1 yields a small
range of \teq{\mu \approx 1} over which \teq{\hat{Q}} varies rapidly with angle.  This 
character is perhaps best elucidated by inserting the result from Eq.~(\ref{eq:calA_calC_lowfreq}) 
into Eq.~(\ref{eq:hatI_Whitney}), leading to
\begin{equation}
   \hat{I}_\parallel \approx \; \dover{4\bigl(1-\mu^2\bigr)+ (\omega/\wcyc)^2}{4 \bigl( 1 - \mu^2\bigr) + 2(\omega/\wcyc)^2}
   \quad ,\quad
   \hat{I}_\perp \; \approx \; \dover{ (\omega/\wcyc)^2}{4 \bigl( 1 - \mu^2\bigr) + 2(\omega/\wcyc)^2} 
   \quad \Rightarrow\quad
   \hat{Q} \;\approx\; \dover{4\bigl(1-\mu^2\bigr)}{4 \bigl( 1 - \mu^2\bigr) + 2(\omega/\wcyc)^2} \quad .
 \label{eq:hatI_Whitney_lowfreq}
\end{equation}
From these relations, it is clear that when \teq{1-\mu^2 = \sin^2\theta \ll  (\omega/\wcyc)^2}, 
the configuration consists of roughly equal populations of the two linear polarization
modes, \teq{\hat{I}_\parallel \approx 1/2 \approx \hat{I}_\perp}, corresponding to
\teq{\hat{Q}\approx 0}. This is not unexpected since the photon angle 
\teq{\theta = \arccos (\kvechat\cdot\Bvechat )} is very close to the field direction.  
One can thus identify a {\bf magnetic scattering cone}
of opening angle \teq{\thetaB = \omega/\wcyc} around \teq{\Bvec}, within which 
(i.e., for \teq{\theta < \thetaB}) the linear polarization degree is modest or small, and outside 
of which it is essentially 100\% due to the dominance of the \teq{\parallel} mode in the 
equilibrium configuration.  This peculiar character within a small solid angle
around \teq{\Bvec} actually has an important impact upon simulations of 
magnetar atmospheres, as will become apparent in Sec.~\ref{sec:discuss} below.

The next domain of interest is at the cyclotron frequency, i.e. \teq{\omega = \wcyc}, 
approximately sampled by hard X-ray emission from accretion columns in X-ray binaries.
The cross section is resonant there and \teq{\DeltaB \approx \SigmaB \gg 1} 
\citep[technically they are infinite, except when introducing a cyclotron width in 
quantum treatments:][]{HD-1991-ApJ,Gonthier-2014-PhRvD}, 
so that the two factors inside the curly braces in the integrands in Eq.~(\ref{eq:master_system}) 
are approximately equal.  From this we deduce that \teq{1 + {\cal A} = 2 {\cal C}}
at the resonance.  Furthermore, from Eq.~(\ref{eq:sigma_tot_asymptotic})
we find that \teq{\sigma  (\omega ,\, \mu_i )/\sigt \approx \SigmaB \bigl(
{\cal K}(\mu_i) + 2 [ 1+ {\cal A} ] \, \mu_i^2\bigr)\, /\,(1+ {\cal A}\mu_i^2)/2}, much of which directly 
cancels the factors inside the curly braces in the numerators of the two system equations.
Evaluation is now simple:
\begin{equation}
   2{\cal C} \; =\; 1 + {\cal A} \; = \; \dover{3}{2} \int_0^1 \bigl( 1 + {\cal A} \mu_i^2 \bigr) \, d\mu_i
   \; =\; \dover{3}{2} + \dover{\cal A}{2} 
   \quad \Rightarrow \quad
   {\cal A}\; =\; 1 \; =\; {\cal C}\quad \hbox{for} \quad  \omega \; =\; \wcyc\quad . 
 \label{eq:calAeq_res}
\end{equation}
An alternative path for inferring this is to use Eq.~(\ref{eq:solution2}) directly, for which 
the integral simply approaches zero at the cyclotron resonance due to the divergence 
of the cross section, \teq{\sigma  (\omega ,\, \mu_i )/\sigt \approx \SigmaB \bigl( 1 + \mu_i^2 \bigr)}.

The non-magnetic domain \teq{\omega \gg \wcyc} is pertinent to surface X-rays from 
millisecond pulsars, and ultra-violet emission from white dwarf atmospheres of not too high 
a magnetization.   To leading order, the cross section is then just the Thomson one 
\teq{\sigt} (since \teq{\SigmaB\approx 1}), and so only the numerator terms enter into the integrations
in Eq.~(\ref{eq:master_system}).  It then follows that both integrations are routinely 
tractable.  Since the circularity is small in this domain, namely \teq{\DeltaB\ll 1}, terms 
containing it in the numerator contribute only to higher order.
One quickly deduces that both \teq{{\cal A}\ll 1} and \teq{{\cal C}\ll 1}.  The 
\teq{2{\cal C}} master equation then yields
\begin{equation}
   2 \, {\cal C} \; \approx\;  \dover{3}{4} \int_0^1  d\mu_i\, 
   \Bigl\{  \DeltaB \left( 1 + \mu_i^2 \right)   + 4 \SigmaB {\cal C} \, \mu_i^2 \Bigr\}
   \; \approx\; \DeltaB + {\cal C} \quad ,
 \label{eq:soln2_high_freq}
\end{equation}
from which one deduces that \teq{{\cal C} \approx \DeltaB \approx 2\wcyc/\omega}.
Manipulating the \teq{1 + {\cal A}} equation is a bit more involved, as it 
requires including corrections to the cross section \teq{\sigma (\omega ,\, \mu_i)\approx \sigt}
and forming the next order contribution from its Taylor series in the numerator 
of the integrand.  This correction includes terms of order \teq{\SigmaB-1} and 
\teq{\DeltaB}.  The algebra is routine, and the result is an identity for \teq{\cal A}
in terms of \teq{\SigmaB-1} and \teq{\DeltaB {\cal C}}.  These developments yield
\teq{{\cal A} \approx 17 \wcyc^2/(2 \omega^2)}, which is much smaller than 
\teq{\cal C}.  Therefore, the anisotropy and linear polarization content of the 
high opacity radiation field is of smaller order than its circulation polarization.
These results and those for the other two frequency domains are summarized 
in Table~\ref{tab:calA_calC_csect_asymp}.

\begin{table}[h]
\vspace*{-0.0cm}
\begin{center}

\caption{Linearity and Circularity Functions, Parameters and the Cross Section}
   \label{tab:calA_calC_csect_asymp}

\hspace{-45pt}\begin{tabular}{l c c c c c }
\hline
\hline\\[-10pt]
  & ${\cal A}$ & ${\cal C}$ & $\SigmaB$ & $\DeltaB$ & $\;\;\sigma (\omega ,\, \mu )/\sigt\;\;$ \\[2pt]
\hline
\hline\\[-9pt]
$\omega\ll\wcyc$ &          $ -1 + \dover{1}{2}\,  \left( \dover{\omega}{\wcyc} \right)^2$ 
                & $ \dover{1}{2}\,  \left( \dover{\omega}{\wcyc} \right)^3$
                & $ \left( \dover{\omega}{\wcyc} \right)^2$ 
                & $ 2 \left( \dover{\omega}{\wcyc} \right)^3$ 
                & $1-\mu^2$ \\[10pt]
$\omega\approx\wcyc$  & $1$ & $1$ & $\gg 1$ & $\SigmaB$  
                & $\SigmaB \bigl( 1 + \mu^2 \bigr) $ \\[4pt]
$\omega\gg\wcyc$ &         $ \dover{17}{2}   \left( \dover{\wcyc}{\omega} \right)^2$
                & $2\, \dover{\wcyc}{\omega} $ 
                & $1 + 2 \left( \dover{\wcyc}{\omega} \right)^2$ 
                & $ 2\, \dover{\wcyc}{\omega} $ 
                & $1$ \\[8pt]
\hline
\hline
\end{tabular}
\end{center}
 \label{tab:calA_calC_asymptotics}
\end{table}
\vspace{-0pt}

The last focus here is on the special frequency \teq{\omega = \wcyc/\sqrt{3}}, 
at which the total cross sections for the two standard linear polarization 
states \teq{\perp} and \teq{\parallel} (relative to the \teq{\kvec - \Bvec} plane)
both coincide with the Thomson value, and are independent of the incoming 
photon direction: see Eq.~(B4) and Figure~B1 of \cite{Barchas-2021-MNRAS}.
This circumstance constitutes a ``linear mode collapse'' in that the circular 
polarization mode is just as prominent as either of the linear eigenmodes.  
This is also reflected in the full polarization description of the cross section 
in Eq.~(\ref{eq:sigma_tot_Stokes}), wherein \teq{\SigmaB=1} and \teq{\DeltaB = \sqrt{3}/2}.  
When \teq{\omega = \wcyc/\sqrt{3}}, the cross section reduces to 
\teq{\sigma (\omega, \, \mu ) = \sigt [1 + ({\cal A} + {\cal C}\sqrt{3})\mu^2] / (1 + {\cal A} \mu^2)},
and in the neighborhood of this frequency, a series expansion leads 
to the approximation
\begin{equation}
   \bigl( 1 + {\cal A}\mu^2 \bigr) \, \dover{\sigma (\omega ,\, \mu )}{\sigt} 
   \; \approx\; \Bigl[ 1 + ({\cal A} + {\cal C}\sqrt{3})\mu^2 \Bigr] 
   \; +\; \bigr( \SigmaB - 1 \bigr) \biggl\{ \bigl( 1 + {\cal A}\mu^2 \bigr) 
     -  \bigl( 1- {\cal A} + 2 {\cal A} \mu^2 \bigr) \, \dover{ 1- \mu^2}{2} 
    + \dover{10\, {\cal C}}{3\sqrt{3}} \, \mu^2 \biggr\} \quad ,
 \label{eq:sigma_tot_bifurc_nbhd}
\end{equation}
to leading order, wherein \teq{\SigmaB -1 \approx 9 \bigl[ \sqrt{3} \,\omega/\wcyc - 1\bigr] /2 \ll 1}.
With a modicum of algebraic manipulation, it can then be shown that 
at precisely \teq{\omega = \wcyc/\sqrt{3}}, the master 
equations in Eq.~(\ref{eq:master_system}) are no longer linearly independent, and coalesce into a single integral
equation that cannot be solved for \teq{\cal A} nor \teq{\cal C} independently.  This frequency constitutes 
a bifurcation point in the pathology of the solution space, in the neighborhood 
of which both  \teq{\cal A} and \teq{\cal C} vary comparatively quickly and continuously with 
frequency \teq{\omega}. Numerical solution thus cannot be obtained
at \teq{\omega = \wcyc/\sqrt{3}}, and in practice is acquired routinely via interpolation 
of determinations in its immediate neighborhood.

\subsection{Numerical Solution for Coefficients \teq{\cal A} and \teq{\cal C}}
 \label{sec:numerics} 

The mathematical character of the denominator \teq{\sigma (\omega ,\, \mu_i )}
in the integrands on the right of the master equations in Eq.~(\ref{eq:master_system})
dictates our design of the numerical algorithm.  
For the solution space, there are two branches, divided by the \teq{\SigmaB = 1} frequency, 
i.e., \teq{\omega/\wcyc = 1/\sqrt{3}}, where the cross sections for 
linear and circular polarizations coalesce \citep[see Fig.~B1 of][and associated discussion]{Barchas-2021-MNRAS}.  
At higher frequencies where \teq{\SigmaB > 1}, 
the denominator can be factorized in the form \teq{(\alpha \mu_i^2 + 1 )(\beta \mu_i^2 + 1 )},
with both \teq{\alpha} and \teq{\beta} being positive.  This circumstance can be inferred from 
Eq.~(\ref{eq:sigma_tot_bifurc_nbhd}) by setting \teq{\mu\to \mu_i},
and for \teq{\alpha \approx {\cal A} + {\cal C}\sqrt{3}} and \teq{\beta \approx 
\bigl[ \SigmaB - 1 \bigr] {\cal A}/\bigl( {\cal A} + {\cal C}\sqrt{3} \bigr)},
with \teq{\vert\beta\vert \ll \alpha}.  One then observes the positive nature of the 
\teq{\alpha} and \teq{\beta} coefficients, realized since \teq{{\cal A}, {\cal C} > 0}
in this frequency neighborhood.  Thus the analytic evaluation 
of the integrals using partial fractions generates two \teq{\arctan} functions with 
different arguments that are complicated forms involving \teq{{\cal A}, {\cal C}, 
\SigmaB, \DeltaB}.   For low frequencies, below \teq{\omega/\wcyc = 1/\sqrt{3}}, one of the 
\teq{\alpha, \beta} becomes negative and so the pathology of the integrand and 
evaluation changes with one of the \teq{\arctan} functions being replaced by the 
logarithmic (or argtanh) form that serves as an analytic continuation.  In either case, 
the analytic evaluation of the integrals yields complicated transcendental forms for the master 
equations, and so does not facilitate the root solving task, which must still be done 
numerically.  It is actually just as easy to effect this using direct numerical evaluation
of the integrals in Eq.~(\ref{eq:master_system}), and therefore this is the protocol adopted here.

\begin{figure}[htbp]
\begin{center}
\centerline{\includegraphics[width=.48\textwidth]{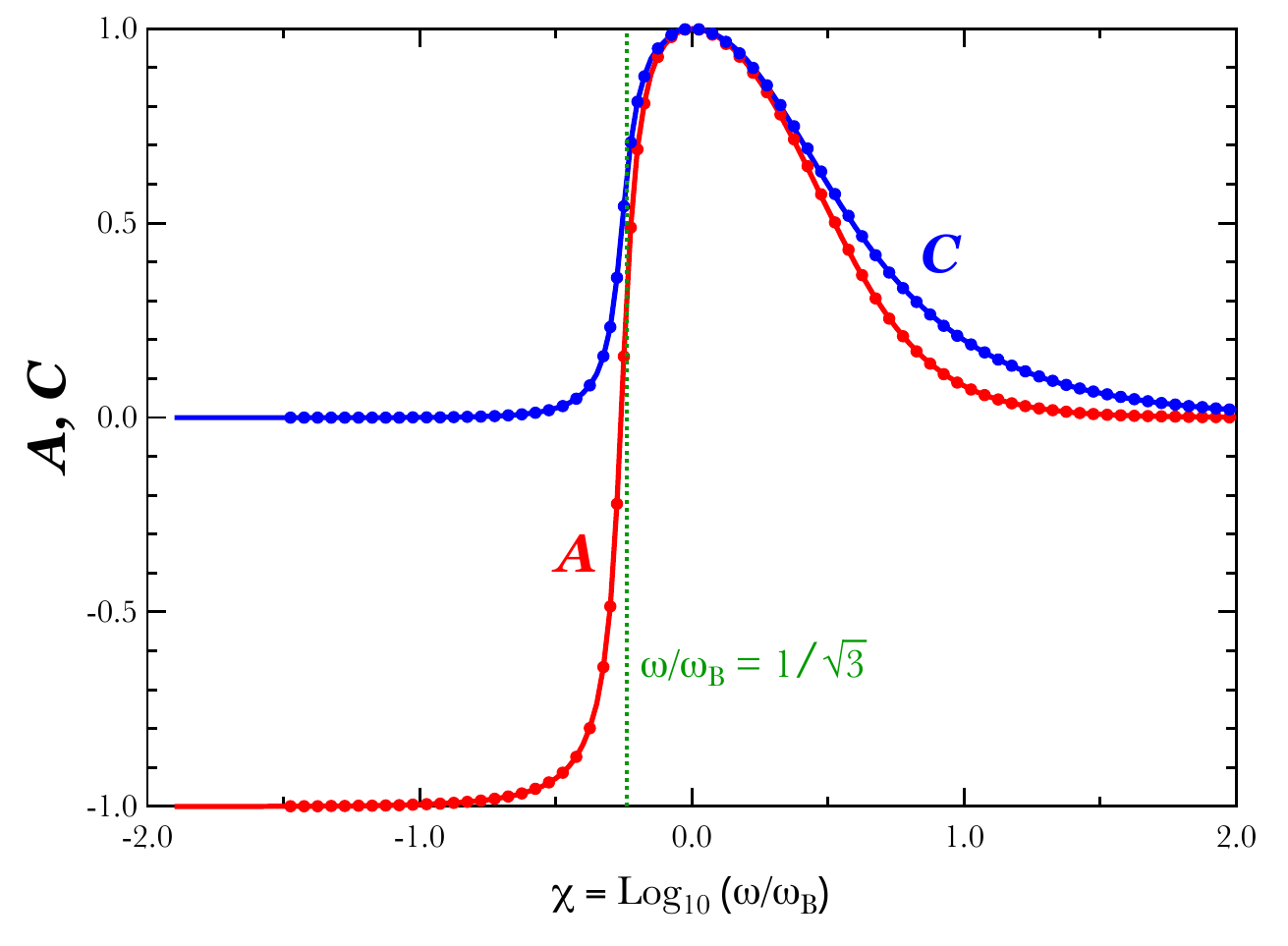}
\hspace{0pt}
\includegraphics[width=.48\textwidth]{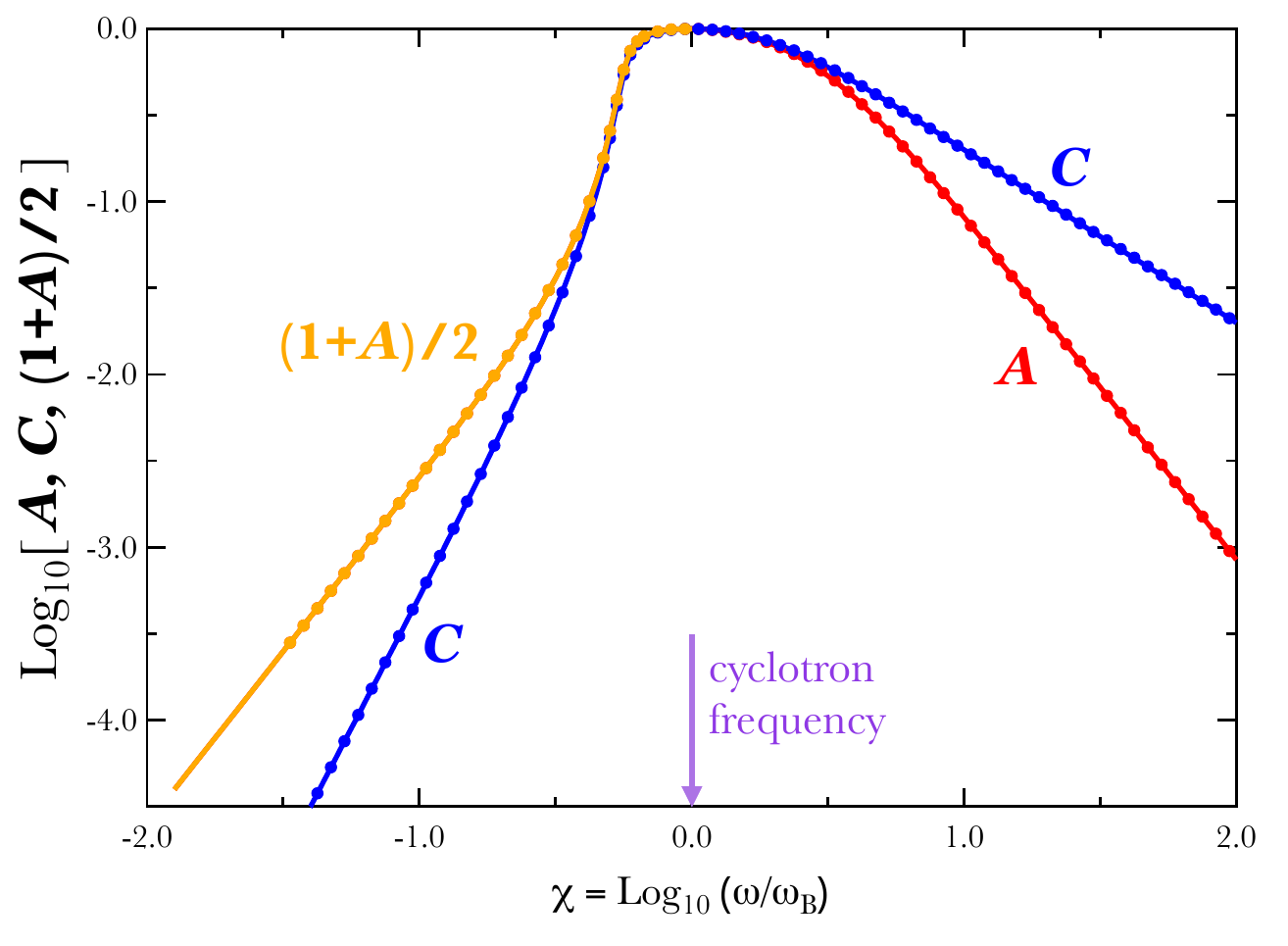}}
\vspace{-10pt} 
\caption{{\it Left}: Numerical solution of the master equations (points) 
at different logarithmic frequencies \teq{\chi = \log_{10} (\omega/\wcyc )} for 
\teq{\cal A} (anisotropy and linear polarization; red) and \teq{\cal C} (circular polarization; blue) 
together with the empirical approximations in Eqs.~(\ref{eq:calA_calCupp_rat_fit}) 
and~(\ref{eq:calA_calClow_rat_fit}) as the solid curves.  
The critical frequency \teq{\omega = \wcyc/\sqrt{3}} where the cross sections for 
linear and circular polarizations are equal, is marked by the vertical green dashed line. 
{\it Right}: A logarithmic version of these solutions that helps visually
illustrate the impressive agreement of the empirical approximations to the 
numerical solution data. While \teq{\cal C} is again represented by the blue curve and dots, 
and the red curve and dots constitute \teq{\cal A} when \teq{\omega \geq \wcyc},  
the \teq{\cal A} information is presented instead as \teq{(1+{\cal A})/2} (orange 
dots and curve) below the cyclotron frequency (marked in mauve).  Both the numerical solutions and 
the approximations accurately match the asymptotic results in Table~\ref{tab:calA_calC_csect_asymp}
in the \teq{\omega \ll \wcyc} and \teq{\omega \gg \wcyc} domains.} 
 \label{fig:calA_calC}
\end{center}
\vspace{-30pt} 
\end{figure}

The numerical algorithm is a two-dimensional root solving problem that presents 
no significant issues.  The variation of the integrals is well-behaved even if not 
monotonic in \teq{{\cal A}} and \teq{{\cal C}}, and so the doublet of equations can be 
solved using a 2D Newton-Raphson technique on the box 
\teq{\vert {\cal A}\vert \leq 1, \; 0 < {\cal C} \leq 1}.  This was performed using 
{\tt Mathematica} code, including the evaluation of the integrals.
Solutions were routinely obtained using different initial values 
\teq{{\cal A}_0} and \teq{{\cal C}_0} in the different \teq{\omega/\wcyc} domains, and the final 
solutions were demonstrated to be insensitive to the choice of the trial 
\teq{{\cal A}_0} and \teq{{\cal C}_0}.  The solutions 
for \teq{{\cal A}} and \teq{{\cal C}} were then fed back into Eq.~(\ref{eq:master_system}),
always yielding identity between left and right hand sides at better than the 
\teq{3\times 10^{-7}} level, and usually orders of magnitude more precise.
Moreover, the numerical solutions reproduced all the low and high frequency 
asymptotic results identified in Table~\ref{tab:calA_calC_csect_asymp} to impressive accuracy,
and are also consistent within a few percent with the numerical solutions obtained from the 
Monte Carlo simulation approach discussed in \cite{Barchas-2021-MNRAS}.
Accordingly, our complete confidence in the robustness of the numerical solutions was established.
A discrete selection of solution values for \teq{{\cal A}} and \teq{{\cal C}} 
is plotted as a function of logarithmic frequency \teq{\chi = \log_{10}(\omega/\wcyc )}
in Fig.~\ref{fig:calA_calC}.  Both coefficients are monotonic either side of their peaks 
at \teq{\omega = \wcyc},  with \teq{{\cal A} \leq {\cal C}} always.

For general utility of results, we opted not to generate extensive tables for display 
but rather to develop relatively compact analytic approximations that are easily 
encoded.  This was an approach adopted in the Monte Carlo simulation analysis 
of \cite{Barchas-2021-MNRAS}, specifying empirical functions for \teq{{\cal A}} and 
\teq{{\cal C}} over the restricted frequency range \teq{0.1 \leq \omega/\wcyc \leq 10}.  
The approximations identified in Eqs.~(26) and~(29) of \cite{Barchas-2021-MNRAS}
were of complex mathematical pathology that did not capture the polynomial character 
of the asymptotics listed in Table~\ref{tab:calA_calC_csect_asymp}.  Accordingly, they 
generally degraded in accuracy outside their frequency domain of development. 
Here, we deliver improved empirical approximations that are ratios of polynomials
in the scaled frequency variable \teq{x=\omega/\wcyc}; we opt to use the \teq{x} variable 
mostly hereafter to render the ensuing mathematical expressions more compact.  
As in \cite{Barchas-2021-MNRAS},  it proved necessary to divide the frequency space 
into \teq{x<1} and \teq{x\geq 1} domains, and derive separate fits.  A least squares 
fitting tool in {\tt Mathematica} was employed, and the degrees of the polynomials 
in the numerator and the denominator were increased until an acceptable fit was achieved.  
The approximant was constrained by the \teq{x\ll1} and \teq{x\gg 1} asymptotic forms in 
Table~\ref{tab:calA_calC_csect_asymp}, as appropriate, and it was also constrained 
so that its first derivative in \teq{x} was zero at \teq{x=1}, where both \teq{{\cal A}} and 
\teq{{\cal C}} achieve local maxima and values of unity.  The resulting forms for the 
\teq{x\geq 1} range are
\begin{equation}
   {\cal A} (x ) \;\approx\; \dover{ 17 x^4 + 297 x^2 -30}{x^2 (2 x^4 + 43 x^2 + 239)} 
   \quad ,\quad
   {\cal C} (x)\;\approx\; \dover{2x^2 + 29 x +15}{x^3 + 15 x^2 + 30} 
   \quad ,\qquad x\; =\; \dover{\omega}{\wcyc} \;\geq\; 1\quad .
 \label{eq:calA_calCupp_rat_fit}
\end{equation}
The accuracies of these approximations are better than 0.8\% for \teq{\cal A}
and better than 0.9\% for \teq{\cal C} at all frequencies.

For the \teq{x = \omega/\wcyc \leq 1} domain, the empirical approximation fitting is somewhat 
more involved because of the functional dependences.  The \teq{{\cal A},{\cal C}} 
variations in \teq{x} have a sharper transition between domains of shallow gradient, to steep 
gradient, to shallow gradient.  This forces the inclusion of around double the number 
of terms in rational function approximations in order to deliver comparable
precision.  In addressing this \teq{x\leq 1} domain, it is simpler to work with 
\teq{{\cal A}' \equiv (1+{\cal A})/2} as a proxy for \teq{\cal A} since it has a cleaner 
power-law character when \teq{x\ll 1}: see Table~\ref{tab:calA_calC_csect_asymp}.
Furthermore, both \teq{{\cal A}'} and \teq{\cal C} posses peak values of unity 
at the cyclotron frequency (\teq{x=1}), and therefore have zero derivative in \teq{x} there.  This 
character constrains the forms that the approximations can take.  After considerable 
exploration, it was found that it was convenient to express the sub-cyclotronic 
rational function approximations using the proxy variable \teq{ \lambda (x) =1/x - x}
to force derivatives to zero at \teq{x=1} where \teq{\lambda (x)=0}.  For \teq{x < 1}, 
we have \teq{\lambda (x) > 0}. The empirical fitting functions then assumed the forms
\begin{eqnarray}
  {\cal A}' & \equiv & \dover{1+{\cal A}}{2}  \; \approx \;
   \Biggl\{ 1 + \sum_{j=1}^4 n_j^{\cal A} \bigl[ \lambda (x) \bigr]^{2j} \Biggr\} \Biggl/ 
   \Biggl\{ 1 + \sum_{j=1}^5 d_j^{\cal A} \bigl[ \lambda (x) \bigr]^{2j} \Biggr\} \nonumber\\[-5.5pt]
 \label{eq:calA_calClow_rat_fit} \\[-5.5pt]
   {\cal C} & \approx & \dover{1}{1+ 2 \lambda^3 (x)}  
   \Biggl\{ 1 + \sum_{j=2}^5 n_j^{\cal C} \bigl[ \lambda (x) \bigr]^{2j} \Biggr\} \Biggl/ 
   \Biggl\{ 1 + \sum_{j=1}^5 d_j^{\cal C} \bigl[ \lambda (x) \bigr]^{2j} \Biggr\} 
   \quad \hbox{for}\quad
   \lambda (x) \; =\; \dover{1}{x} - x  \quad .\nonumber
\end{eqnarray}
These forms automatically match \teq{{\cal A}=1={\cal C}} at \teq{x=1}, and 
have zero derivatives there; they have Taylor series quadratic in 
\teq{(x-1/x)} about \teq{x=1}.  To match the \teq{{\cal A}'\approx x^2/4} 
asymptotic limit for \teq{x\ll 1}, we set \teq{n_4^{\cal A}=d_5^{\cal A}/4}. 
For \teq{\cal C},  when \teq{x\ll 1}, the correct asymptotic behaviour is obtained if 
\teq{n_5^{\cal C}=d_5^{\cal C}}.  The least squares fitting protocol delivered numerical
values for the numerator \teq{n_j^{{\cal A},{\cal C}}} and denominator \teq{d_j^{{\cal A},{\cal C}}} 
coefficients, with 8 free parameters.  We judiciously approximated these by 
rational numbers without any significant degradation of precision, and in fact a 
slight improvement of such.  The resulting coefficients are listed in Table~2.
The empirical fits are accurate to better than 
\teq{0.9}\% for both \teq{\cal A} and \teq{\cal C} for all \teq{x\leq 1}, and mostly better.

\begin{table}[h]
\vspace*{-0.0cm}
\begin{center}

\caption{Coefficients for \teq{\cal A} and \teq{\cal C} Empirical Fit Rational Functions}

\hspace{-85pt}\begin{tabular}{l c c c c c c c c c l }
\hline
\hline\\[-10pt]
  & $n_1^{{\cal A},{\cal C}}$ & $n_2^{{\cal A},{\cal C}}$ & $n_3^{{\cal A},{\cal C}}$ & $n_4^{{\cal A},{\cal C}}$  & $n_5^{{\cal A},{\cal C}}$
  & $d_1^{{\cal A},{\cal C}}$ & $d_2^{{\cal A},{\cal C}}$ & $d_3^{{\cal A},{\cal C}}$ & $d_4^{{\cal A},{\cal C}}$  & $d_5^{{\cal A},{\cal C}}$\\[2pt]
\hline
\hline\\[-9pt]
${\cal A}'$ & $-\dover{15}{19}$ & $\dover{9}{32}$ & $- \dover{1}{23}$ & $\dover{1}{33}$ &  & 
                    $- \dover{7}{10}$ & $\dover{1}{5}$ & $\dover{1}{5}$ & $- \dover{4}{17} $ & $\dover{4}{33} = 4 n_4^{\cal A}$ \\[10pt]
${\cal C}$ & & $\dover{24}{7}$ & $-\dover{21}{16}$ & $- \dover{8}{23}$ & $\dover{17}{26}$ & 
                    $- \dover{1}{4}$ & $\dover{16}{19}$ & $\dover{2}{7}$ & $- \dover{15}{14}$ & $ \dover{17}{26} = n_5^{\cal C}$ \\[8pt]
\hline
\hline
\end{tabular}
\end{center}
 \label{tab:calA_calC_empir_coeff} 
\vspace{-10pt} 
\end{table}

\subsection{The Character of the Radiation Anisotropy and Polarization}.
 \label{sec:anis_pol_char} 

\begin{figure}[htbp]
\begin{center}
\centerline{\includegraphics[width=.49\textwidth]{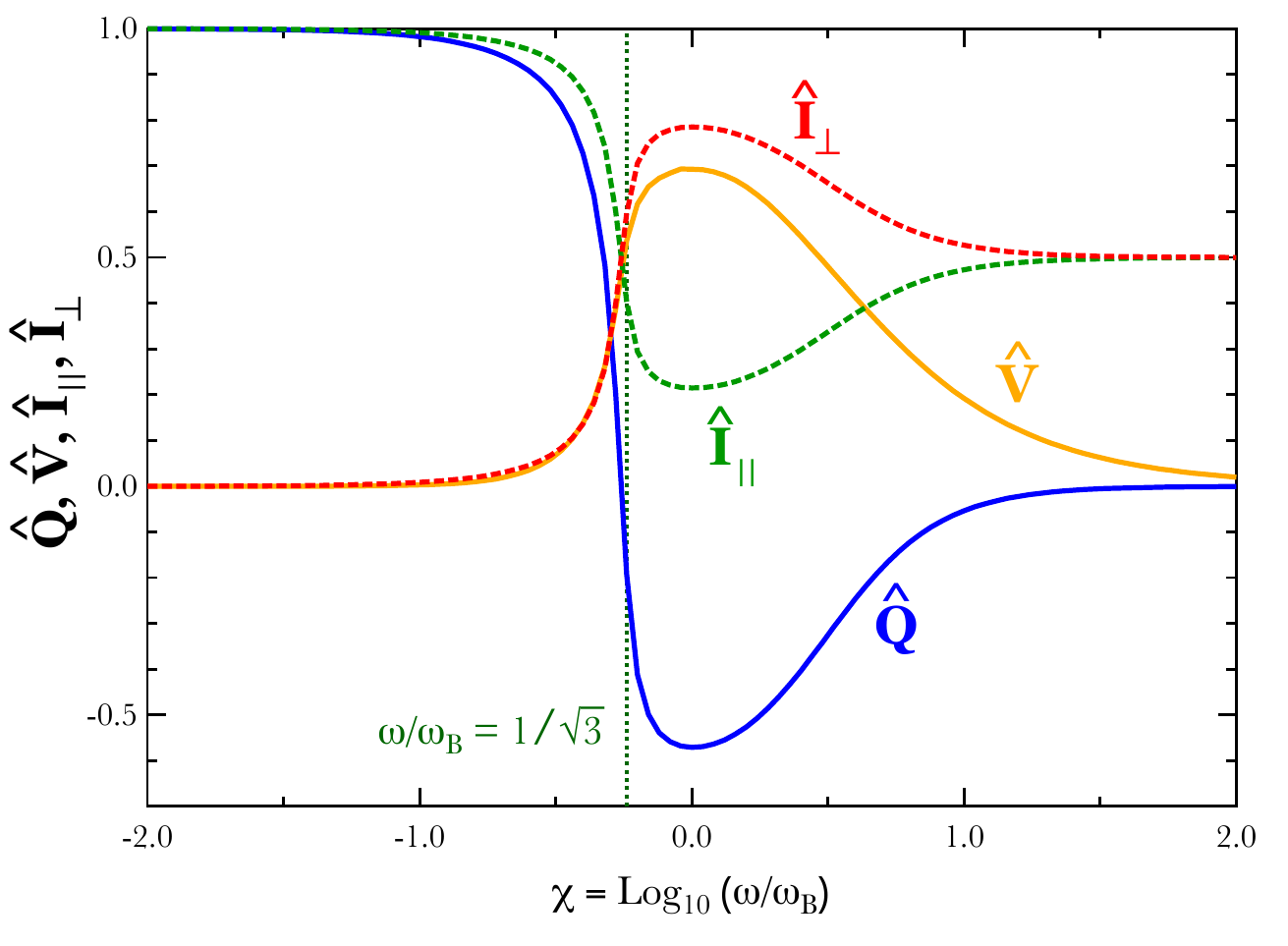}
     \hspace{0pt} \includegraphics[width=.49\textwidth]{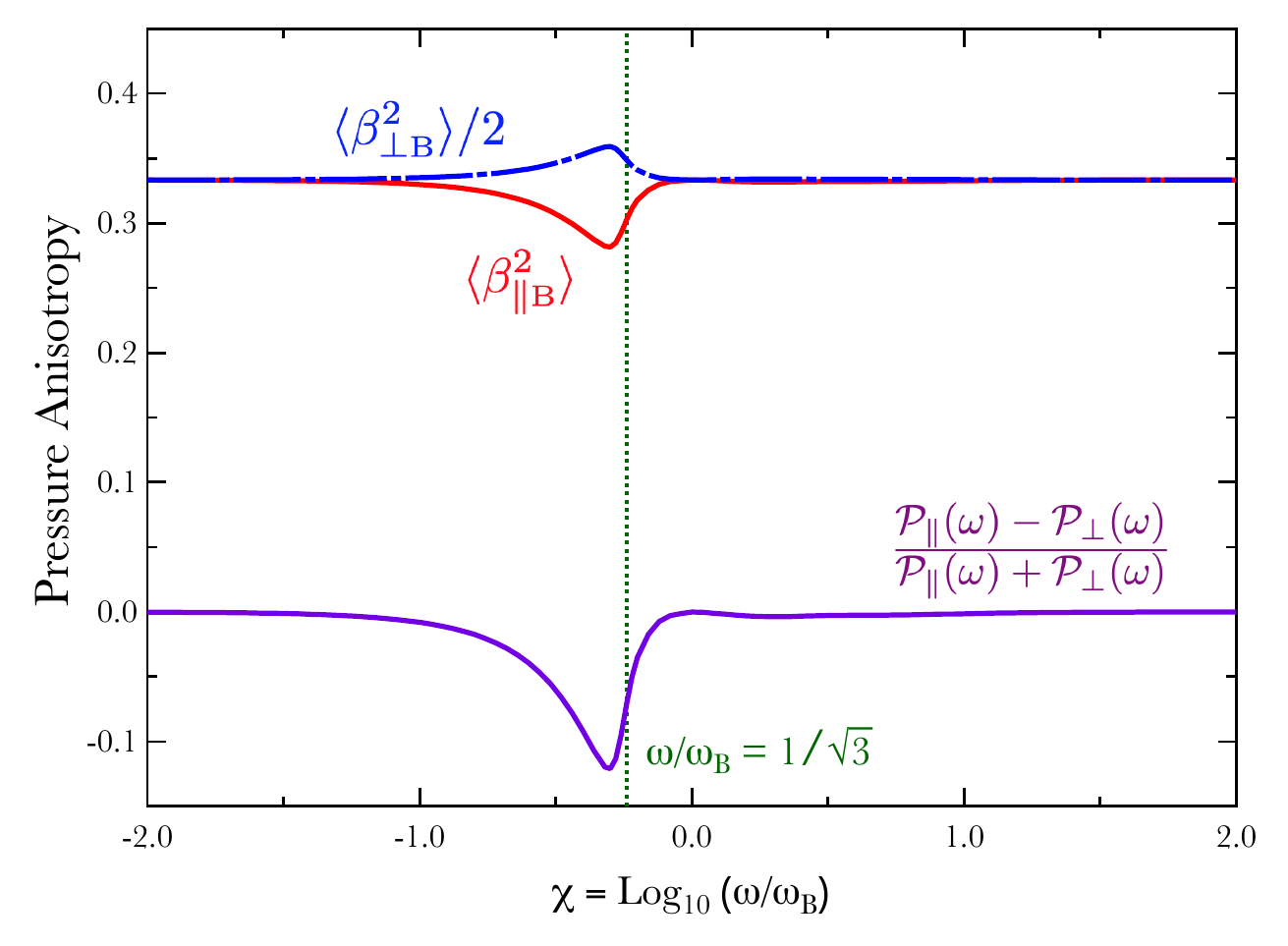} } 
\vspace{-8pt}
\caption{{\it Left}: Angle-averaged polarization measures \teq{\bigl\langle \hat{Q} \bigr\rangle} 
and \teq{\bigl\langle \hat{V} \bigr\rangle}, as defined in Eq.~(\ref{eq:hatQ_hatV_ave}), and 
\teq{\bigl\langle \hat{I}_{\parallel} \bigr\rangle} and \teq{\bigl\langle \hat{I}_{\perp} \bigr\rangle}, 
as given in Eq.~(\ref{eq:hatI_par_perp_ave}), as functions of the logarithm of the dimensionless 
photon frequency \teq{\omega/\wcyc}.  In the non-magnetic domain, \teq{\omega\gg\wcyc}, 
they evince an unpolarized configuration. In the highly-magnetic regime, \teq{\omega\ll\wcyc}, 
dominance of the \teq{\parallel} polarization mode (\teq{Q\to 1}) is clear.  All quantities exhibit 
extrema at the cyclotron frequency, and vary rapidly around \teq{\omega/\wcyc = 1/\sqrt{3}}.
{\it Right}: Radiation anisotropy measures \teq{\langle\betaparsq\rangle}, as defined in 
Eq.~(\ref{eq:betaparsq_eval}), and its counterpart \teq{\langle\betaperpsq\rangle/2}.  These 
two are approximately equal to \teq{1/3} when \teq{\omega\gtrsim \wcyc} and also when 
\teq{\omega\ll \wcyc}, when approximate isotropy exists; only when \teq{\omega/\wcyc \sim1/\sqrt{3}} 
is there significant radiation anisotropy.  Also depicted is the pressure anisotropy function 
given in Eq.~(\ref{eq:press_anis}).  A modest dominance of pressure perpendicular to \teq{\Bvec} at 
sub-cyclotronic \teq{\omega/\wcyc \sim 1/\sqrt{3}} frequencies is evident. 
}
 \label{fig:hatQV_Ipar_Iperp_press}
\end{center}
\vspace{-25pt} 
\end{figure}

The general character of the anisotropy and the polarization of the high opacity
radiation configuration can be appraised by computing averages of the pertinent 
quantities over the pre-scattering photon angles.  To this end, we now form integrals 
of \teq{\hat{Q}}, \teq{\hat{V}} in Eq.~(\ref{eq:hatIQV}) over \teq{\mu} on the interval \teq{0\leq\mu\leq 1}, 
which serve as average measures.  Note that other averages are possible, for 
example forming integrals over \teq{Q} and \teq{V} divided by that for \teq{I}; adopting such
does not materially alter the conclusions we draw just below.  The averages resulting 
from our protocol are
\begin{equation}
   \bigl\langle \hat{Q} \bigr\rangle \; =\; 1 - \bigl( 1 + {\cal A} \bigr) u \bigl( {\cal A} \bigr)
   \quad ,\quad
   \bigl\langle \hat{V} \bigr\rangle \; =\;  \dover{{\cal C}\, \log_e \bigl( 1 + {\cal A} \bigr) }{{\cal A}} 
    \;\quad \hbox{for}\quad\;
    u (z) \; =\; 
   \begin{cases}
       \dover{\arctan (\sqrt{z})}{\sqrt{z}} 
        \quad , & \text{if}\;\; z > 0 \;\; ,\\
       \dover{1}{2\sqrt{\vert z\vert}} \log_e \dover{1+\sqrt{\vert z\vert }}{1-\sqrt{\vert z\vert}}  
       \quad , & \text{if}\;\; z < 0 \;\; . \\
    \end{cases} 
 \label{eq:hatQ_hatV_ave}
\end{equation}
Similar forms can quickly be obtained for the accompanying linear polarization quantities 
\teq{\hat{I}_{\parallel}} and \teq{\hat{I}_{\perp}}:
\begin{equation}
   \bigl\langle \hat{I}_{\parallel} \bigr\rangle \; =\; 1 - \bigl( 1 + {\cal A} \bigr) \dover{u \bigl( {\cal A} \bigr)}{2} 
   \quad ,\quad
   \bigl\langle \hat{I}_{\perp} \bigr\rangle \; =\; \bigl( 1 + {\cal A} \bigr) \dover{u \bigl( {\cal A} \bigr)}{2} 
   \; =\; 1 - \bigl\langle \hat{I}_{\parallel} \bigr\rangle \quad .
 \label{eq:hatI_par_perp_ave}
\end{equation}
These forms are now purely functions of the frequency ratio, \teq{x=\omega/\wcyc}, and are 
plotted in Fig.~\ref{fig:hatQV_Ipar_Iperp_press}, left, to illustrate the dependence on \teq{x}. 
A number of features are apparent, all being consequences of the character of the 
scattering cross section that is presented in Appendix B of \cite{Barchas-2021-MNRAS}.
In the low frequency domain, \teq{\omega\ll \wcyc}, there is a clear dominance of the \teq{\parallel} 
polarization mode, with \teq{\bigl\langle \hat{I}_{\parallel} \bigr\rangle \to 1} corresponding to \teq{Q\to 1}.
There are rapid variations in all these average polarization quantities in the neighborhood
of \teq{\omega/\wcyc = 1/\sqrt{3}} as the competition between circular and linear polarization 
in the scattering cross section becomes significant.  Circular polarization peaks at \teq{\omega = \wcyc},
with \teq{\bigl\langle \hat{V} \bigr\rangle = \log_e2 \approx 0.69}, accompanied by a minimum 
of \teq{\bigl\langle \hat{Q} \bigr\rangle = 1 - \pi/2 \approx -0.57}.  Around the cyclotron frequency and above it, 
the \teq{\perp} mode is dominant in the photon configuration.  Finally, in the non-magnetic regime, 
\teq{\omega\gg\wcyc}, the high opacity configuration is essentially unpolarized, with
\teq{\bigl\langle \hat{I}_{\parallel} \bigr\rangle \approx \bigl\langle \hat{I}_{\perp} \bigr\rangle \approx 1/2}.

An additional piece of information concerns the anisotropy of the radiation field.
By inspection of Eq.~(\ref{eq:calNALC_def}), the angular intensity is proportional
to \teq{(1+ {\cal A}\mu^2)/\sigma (\omega ,\, \mu )}, where \teq{\sigma (\omega ,\, \mu )} 
is expressed in Eq.~(\ref{eq:sigma_tot_asymptotic}).  This connection with the \teq{\Lambda_I(\mu )} 
redistribution anisotropy function is discussed at length in Section 5.3 of \cite{Barchas-2021-MNRAS}.
There are two anisotropy measures that connect to the pressure that the radiation 
field can exert on a strongly-magnetized electron gas along the direction of \teq{\Bvec},
and perpendicular to it.  The pertinence of radiation pressure is discussed briefly 
in Sec.~\ref{sec:discuss}.  The pressure tensor consists of summations over \teq{p^iv^j} 
products from components of momentum (\teq{p^i}) in the \teq{i^{th}} coordinate direction, 
and velocity (\teq{v^j}) in the \teq{j^{th}} direction:
\begin{equation}
   {\cal P}_{ij} \; =\; \int  n_{\gamma} \bigl( \mathbf{p} \bigr) \, p^iv^j \, d^3p \quad ,
 \label{eq:P_ij_def}
\end{equation}
where \teq{n_{\gamma}\bigl( \mathbf{p} \bigr) \propto (1+ {\cal A}\mu^2)/\sigma (\omega ,\, \mu )} 
is the number density of photons per momentum interval \teq{\mathbf{dp}} that 
can be obtained directly from the angle-dependent intensity.  Clearly, \teq{{\cal P}_{ij}} 
constitutes the purely spatial portion of the energy-momentum tensor \citep{Weinberg-1972-book} 
in the special case where there are no bulk relativistic motions.  Considering only diagonal elements, 
the two pertinent radiation anisotropy measures can be distilled into averages of the 
velocity components (in units of \teq{c}) parallel to (\teq{\betaparsq = \mu^2}) and orthogonal 
(\teq{\betaperpsq = 1-\mu^2}) to \teq{\Bvec}, and are obviously not independent quantities.  
The anisotropic pressure tensor diagonal elements are thus  
\teq{{\cal P}_{\parallel}(\omega)\propto \langle\betaparsq\rangle} along \teq{\Bvec}, and 
\teq{{\cal P}_{\perp}(\omega)\propto \langle\betaperpsq\rangle/2 = \bigl[ 1 - \langle\betaparsq\rangle\bigr]/2} 
in the plane that is orthogonal to the field, noting that the coefficients of proportionality 
in these relations are identical.  If one scales the cross section according to 
\teq{\sigma (\omega ,\, \mu )/\sigt = \overline{\sigma}(\omega ,\, \mu )/ ( 1 + {\cal A}\mu^2 )}, 
so that \teq{\overline{\sigma}(\omega ,\, \mu )} is a dimensionless quadratic function of \teq{\mu^2}, 
one can form the average 
\begin{equation}
   \langle\betaparsq\rangle \; =\; 
   \int_0^1 \dover{\mu^2 \bigl( 1 + {\cal A} \mu^2 \bigr)^2}{\overline{\sigma}(\omega ,\, \mu )} \, d\mu 
   \Biggl/ \int_0^1 \dover{ \bigl( 1 + {\cal A} \mu^2 \bigr)^2}{\overline{\sigma}(\omega ,\, \mu )} \, d\mu \quad .
 \label{eq:betaparsq_eval}
\end{equation}
The pressure anisotropy at a single photon frequency is then 
\begin{equation}
   \dover{{\cal P}_{\parallel}(\omega) - {\cal P}_{\perp}(\omega)}{{\cal P}_{\parallel}(\omega) + {\cal P}_{\perp}(\omega)}
   \; =\; \dover{3 \langle\betaparsq\rangle -1}{\langle\betaparsq\rangle +1} \quad .
 \label{eq:press_anis}
\end{equation}
This pressure anisotropy function of \teq{\omega/\wcyc} is plotted along with 
\teq{\langle\betaparsq\rangle} and \teq{\langle\betaperpsq\rangle/2} in the right 
panel of Fig.~\ref{fig:hatQV_Ipar_Iperp_press}.  At \teq{\omega\ll \wcyc} frequencies, and also
in the non-magnetic \teq{\omega/\wcyc \gg 1} 
domain, \teq{ \langle\betaperpsq\rangle/2 \approx \langle\betaparsq\rangle \approx 1/3} 
and pressure isotropy prevails: \teq{{\cal P}_{\perp}(\omega) \approx {\cal P}_{\parallel}(\omega)}.
The same circumstance arises in the cyclotron resonance, \teq{\omega \approx \wcyc}.  
For all three of these domains, the cross section and \teq{\cal A} values in Table~\ref{tab:calA_calC_asymptotics}
can be used to show that \teq{ \langle\betaparsq\rangle \approx 1/3}.
In contrast, when \teq{\omega/\wcyc \sim1/\sqrt{3}}, then
\teq{ \langle\betaperpsq\rangle/2 > 1/3 >  \langle\betaparsq\rangle}, 
and more pressure appears at large angles to \teq{\Bvec} than is generally 
in the direction of the field.  Yet, the pressure anisotropy is still only at the \teq{\sim 10}\% level.
These determinations follow from the prevailing general isotropy of the radiation field 
that is evident in Fig.~8 of \cite{Barchas-2021-MNRAS}, with the largest anisotropy 
being for \teq{\omega/\wcyc \sim1/\sqrt{3}}.
To generate the complete pressure information, these results would need to be integrated 
over the photon spectrum pertinent to a particular astrophysical setting; this task is
the purview of other work.

\begin{figure}[htbp]
\begin{center}
\centerline{\includegraphics[width=.49\textwidth]{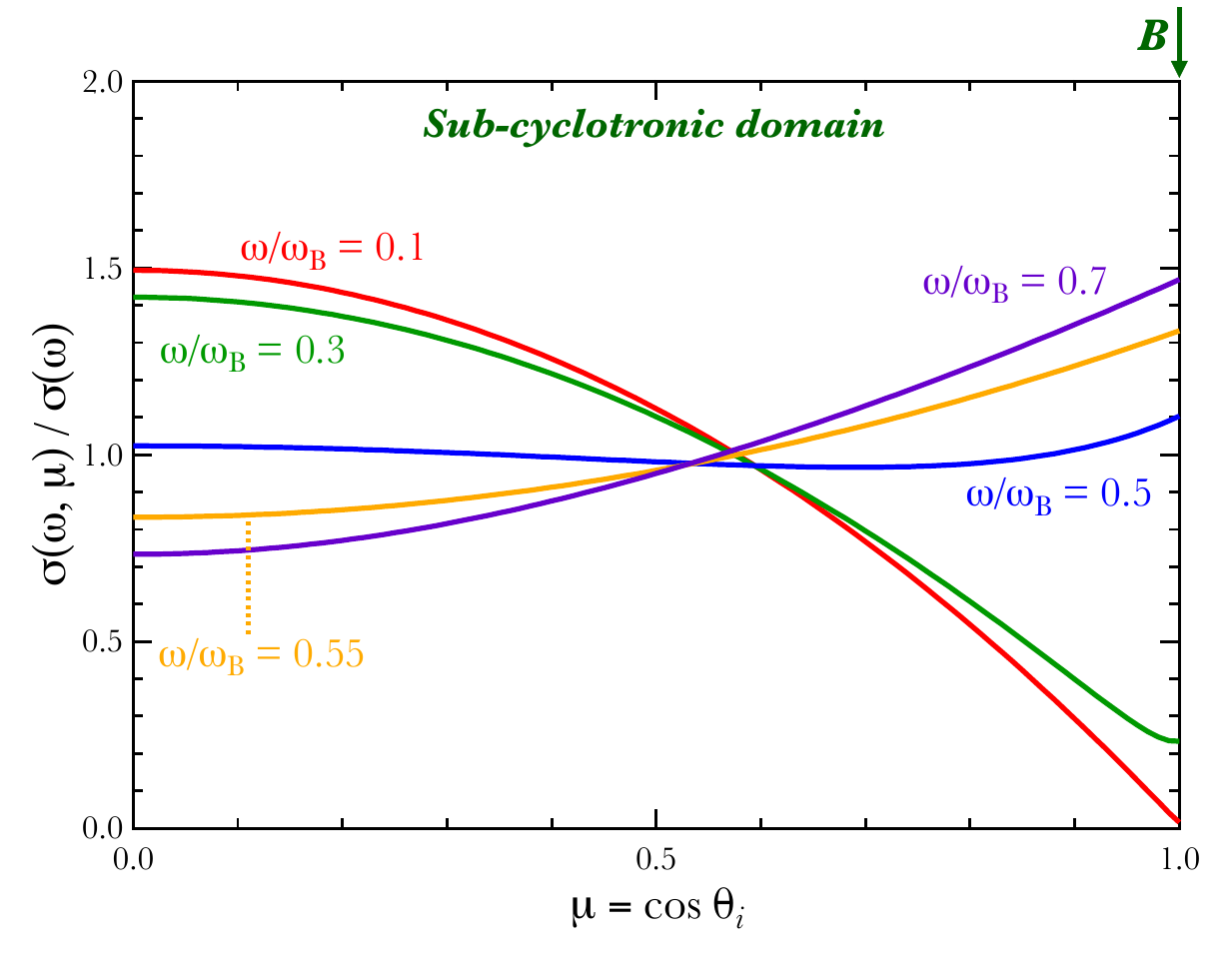}
\hspace{0pt}
\includegraphics[width=.49\textwidth]{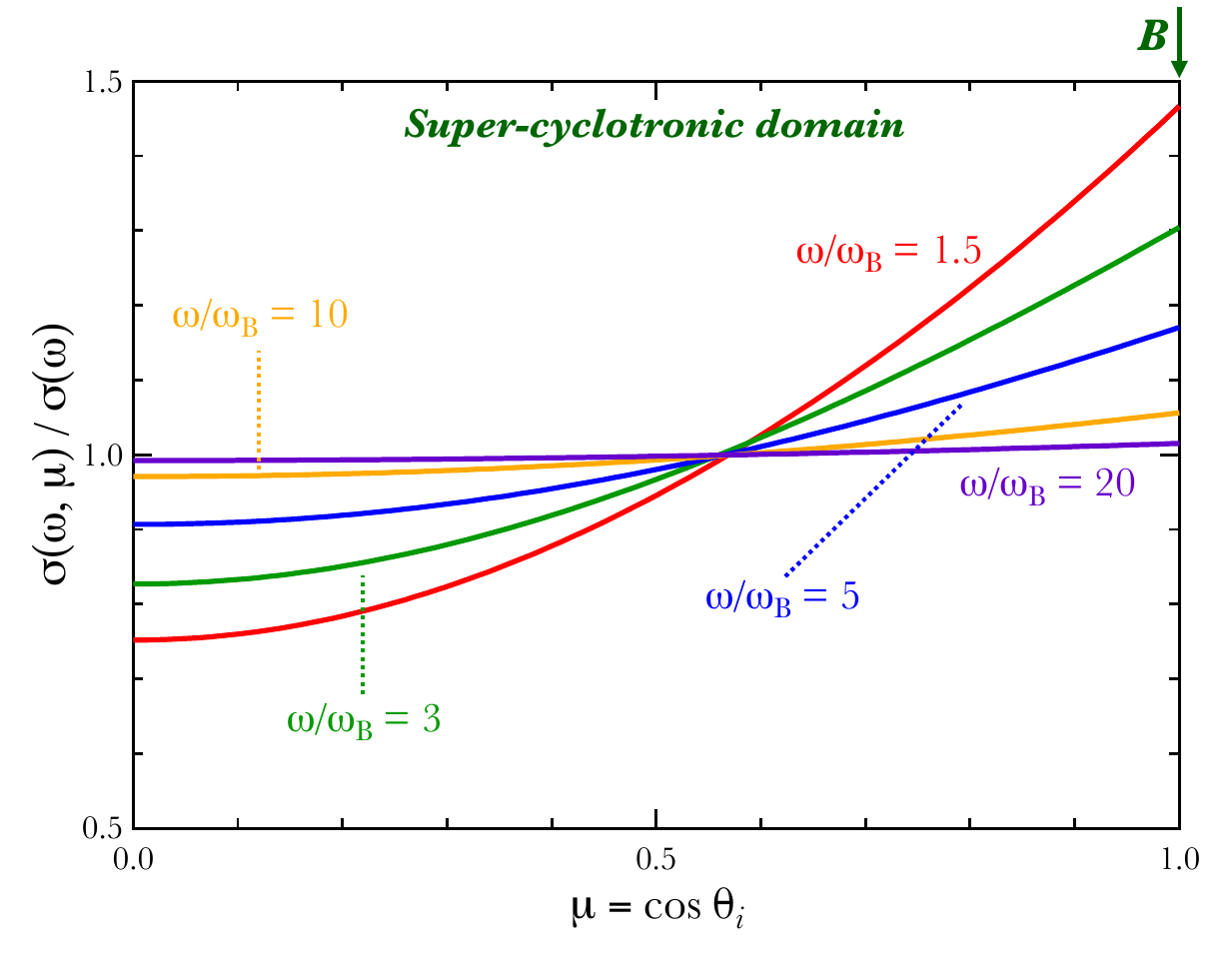}}
\vspace{-8pt}
\caption{The polarized, magnetic Thomson cross section \teq{\sigma (\omega ,\, \mu )}
at high opacities, i.e. that in Eq.~(\ref{eq:sigma_tot_asymptotic}), as 
a function of the angle cosine \teq{\mu = \cos\theta_i} of the incoming photon 
to the magnetic field direction \teq{\boldsymbol{B}} that is indicated at the upper
right of each panel.  The left panel corresponds to the \teq{\omega < \wcyc} 
sub-cyclotron domain, and the right panel is for cases of incoming photons 
above the cyclotron energy.  The polarization configuration is from the empirical formulae in 
Eqs.~(\ref{eq:calA_calCupp_rat_fit}) and~(\ref{eq:calA_calClow_rat_fit}), and the cross sections
are scaled in units of the total (polarized and magnetic) cross section \teq{\sigma (\omega )}. 
When \teq{\omega\gg \wcyc}, the cross section becomes independent of \teq{\mu} and 
\teq{\sigma (\omega ,\, \mu )\to \sigt} as well as \teq{\sigma (\omega )\to \sigt}, 
the familiar non-magnetic regime.  For the \teq{\omega\ll \wcyc} domain, the 
\teq{\omega /\wcyc = 0.1} curve closely approximates the \teq{3(1-\mu^2)/2} 
asymptotic result.  See Table.~\ref{tab:calA_calC_csect_asymp}. }
 \label{fig:magThoms_csect_pol}
\end{center}
\vspace{-20pt} 
\end{figure}

The numerical solution for the coefficients \teq{\cal A} and \teq{\cal C} also enables 
the complete specification of the polarized total cross section \teq{\sigma (\omega,\, \mu)} in 
Eq.~(\ref{eq:sigma_tot_asymptotic}) appropriate for the high opacity domain.
Its behavior is plotted as a function of \teq{\mu} for different frequencies 
in Fig.~\ref{fig:magThoms_csect_pol}, wherein the curves are normalized 
to unit area in the \teq{\mu} variable.  Specifically,  
the cross section can be averaged over all incoming photon angles thus:
\begin{equation}
    \sigma (\omega ) \;\equiv\; \bigl\langle \sigma (\omega ,\, \mu ) \bigr\rangle 
    \; =\; \int_0^1 \sigma (\omega ,\, \mu ) \, d\mu  \quad  ,
 \label{eq:sigma_tot_ave_form}
\end{equation}
which, for \teq{u(z)} as defined in Eq.~(\ref{eq:hatQ_hatV_ave}), 
yields an evaluation
\begin{equation}
   \sigma (\omega )  \; =\;  \sigt \biggl\{ \SigmaB  + \dover{1 - \SigmaB}{2 {\cal A}}  \left[ 1 + \dover{7{\cal A}}{3} 
         - \bigl( 1 + {\cal A} \bigr)^2 u \bigl( {\cal A} \bigr) \right]
    +  \dover{2{\cal C} \DeltaB}{ {\cal A}} \, \Bigl[ 1 - u \bigl( {\cal A} \bigr) \Bigr] \biggr\}. \quad .
 \label{eq:sigma_tot_ave}
\end{equation}
This average is employed to normalize the curves in Fig.~\ref{fig:magThoms_csect_pol} in 
presenting the variations of \teq{\sigma (\omega ,\, \mu ) / \bigl\langle \sigma (\omega ,\, \mu ) \bigr\rangle}.
At \teq{\omega\gg \wcyc} frequencies, \teq{\sigma (\omega ,\, \mu ) \approx \sigt} and the non-magnetic 
Thomson domain is realized.  As \teq{\omega} drops towards the cyclotron frequency, 
the cross section preferentially favors scatterings of photons directed more along the 
field direction, a profound influence on radiative transfer at high opacities.  
This behavior continues smoothly through and below the cyclotron frequency
as the depiction moves from the right to the left panels.  Once the frequency 
drops below the bifurcation value, \teq{\wcyc/\sqrt{3}}, the influence of circular polarizations  
diminishes and the character transitions to describe a \teq{\omega\ll \wcyc} domain 
where scatterings of photons moving approximately perpendicular to \teq{\boldsymbol{B}} 
are more prevalent than those for photons closer to the field direction.
Finally, note that this cross section behavior contributes to the angular dependence 
of the intensity \teq{I(\mu) \propto (1+ {\cal A}\mu^2)/\sigma (\omega ,\, \mu )};
examples of such intensity distributions are given in Fig.~8 of \cite{Barchas-2021-MNRAS}.

\section{Discussion}
 \label{sec:discuss} 

A principal motivation for our derivation of precision numerical solutions to the 
high opacity domain of the polarized Thomson scattering transport equations in cold, 
magnetized plasmas is that it facilitates improved efficiency of transport simulations. 
This is demonstrated in Fig.~2 of \cite{Hu-2022-ApJ}, wherein intensities and Stokes 
parameter solutions for the Monte Carlo simulation {\sl MAGTHOMSCATT} of radiation 
transfer in slabs of fixed Thomson optical depths are compared for two protocols of 
photon injection at the base of the slab.  The first was a simple injection of isotropic and 
unpolarized photons (designated IU injection) at the base of the simulation slab,
identical for all \teq{\omega /\wcyc}, with the ultimate observer signal being recorded for 
those photons that exit the slab at its opposite, upper planar surface. The alternative was 
for a frequency-dependent anisotropic and polarized (AP) photon injection that employed 
the Stokes parameter and anisotropy information encapsulated in Eq.~(\ref{eq:hatIQV}), 
i.e. pertaining to the high opacity configuration.  \cite{Hu-2022-ApJ} performed a limited
exploration of the relative efficacy of the two injection protocols, finding that 
the AP choice generally required slightly or somewhat smaller effective optical 
depths \teq{\tau_{\rm eff}} to realize convergent solutions that did not vary when 
the slab thickness was further increased.  The definition of  \teq{\tau_{\rm eff}} is 
given in Eq.~(5) of \cite{Hu-2022-ApJ}: 
\begin{equation}
   \taueff \; = \; n_e h\, \sigma_{\rm up}(\theta_i \; = \; \thetaB) 
   \; = \;  \dover{\taut}{2} \Bigl\{ \sin^2\thetaB  
              + \SigmaB (\omega) \left[1+\cos^2\thetaB\right] \Bigr\} \quad .
 \label{eq:tau_eff_def}
\end{equation}
It is posited based on the unpolarized Thomson cross section \teq{\sigma_{\rm up}}
evaluated for pre-scattering photon directions along the local slab zenith, and therefore 
at incidence angle \teq{\theta_i=\thetaB} to the field \teq{\Bvec}.
Clearly, while it couples with the Thomson 
value \teq{\taut = n_e\sigt h} for slabs of thickness \teq{h} and electron number 
density \teq{n_e}, it depends significantly on the photon frequency and the 
orientation of the field direction to the slab's zenith/normal.

Here, we deliver a more detailed and precise assessment of how the AP injection 
protocol compares in efficiency with the IU one in {\sl MAGTHOMSCATT}.  For this exploration,
the empirical forms for \teq{\cal A} and \teq{\cal C} in Eqs.~(\ref{eq:calA_calCupp_rat_fit})
and~(\ref{eq:calA_calClow_rat_fit}) were encoded in the simulation so as to test 
their usefulness.  Another upgrade was to render {\sl MAGTHOMSCATT}
in fully parallelized form so that it was run on Rice University's NOTS cluster.  This 
development opened up access to high statistics runs for \teq{\omega/\wcyc \leq 0.1} 
that are generally slower due to the wide disparity of the cross sections over the 
full range of photon polarizations and propagation directions. \teq{\tau_{\rm eff}} values (denoted \teq{\tau_{\rm eff, AP}} 
and \teq{\tau_{\rm eff, IU}}) and associated run times (\teq{\timeAP} and \teq{\timeIU}) 
required to deliver convergent simulations were obtained for a variety of frequency 
ratios \teq{0.01 \leq \omega/\wcyc \leq 10} spanning the cyclotron resonance region.
Results are presented in Table~\ref{tab:IU_AP_inject_simulation}, where the run time 
ratios \teq{\timeIU / \timeAP} are listed since the actual run times depend on the 
scale and configuration of any machine cluster.  Note that all simulations were performed  in exactly 
the same run protocol (and \teq{10^6} photons) so that relative simulation times could be accurately calibrated.
More details on the run times across the spectrum of atmosphere 
parameters \teq{\thetaB} and \teq{\omega/\wcyc} will be provided in Dinh Thi et al. (in preparation).
The definition of convergence was that the output from the upper slab surface 
of all of the relevant Stokes parameter information (\teq{I, Q, V}) does not change appreciably 
when the effective slab optical depth \teq{\tau_{\rm eff}} was increased above the 
nominal values listed here in Table~\ref{tab:IU_AP_inject_simulation}.  This information 
consisted primarily of the zenith angle distributions like those presented in Figure~2 
of  \cite{Hu-2022-ApJ}.  Select zenith angles \teq{\thetaB} for the magnetic field were 
sampled, encompassing magnetic polar, equatorial and mid-latitude locales of the 
Thomson scattering atmospheric slab on a neutron star surface threaded by a 
magnetic dipole.

\begin{deluxetable}{lr|c c c c c c c }[h]
\tablecaption{Effective optical depths \teq{\tau_{\rm eff, AP}} and \teq{\tau_{\rm eff, IU}}
and run time ratios \teq{\timeIU / \timeAP} for IU and AP injection protocols in
{\sl MAGTHOMSCATT}}
\label{tab:IU_AP_inject_simulation}
\tablewidth{0pt} 
\tablehead{
\colhead{} &\colhead{$\omega / \wcyc$} & \colhead{10} & \colhead{3} &
\colhead{0.99} & \colhead{0.3} & \colhead{0.1}
& \colhead{0.03} & \colhead{0.01}
}
\startdata
{  } & $\tau_{\rm eff, IU}$ & 6  & 6  & 9 &  6& 8 & 8 & 10 \\
{  $\thetaB = 0^{\circ}$} & $\tau_{\rm eff, AP}$ & 6  & 6  & 6 &  6& 6 & 6 & 6\\
{  }& $\timeIU / \timeAP \vspace{1.5pt} $&  1 & 0.91  & 1.18& 0.77 & 1.08 & 1.50 & 2.41 \\ 
\hline
{  } & $\tau_{\rm eff, IU}$ & 6 & 8 & 10& 20& 50 & 400& $>$ 800 \\
$\thetaB = 30^{\circ}$ & $\tau_{\rm eff, AP}$ & 6 & 6 & 6& 15& 30 & 150& 200 \\
{  }& $\timeIU / \timeAP \vspace{1.5pt} $ & 0.96& 1.55& 1.04& 1.56& 1.93& 5.76& $>$ 5.69 \\ 
\hline
{  } & $\tau_{\rm eff, IU}$ & 10 & 8& 20& 30 & 100 & 600&$>$ 800  \\
$\thetaB = 60^{\circ}$& $\tau_{\rm eff, AP}$ & 8 &6& 6& 20 & 30 & 200&200  \\
{  }& $\timeIU / \timeAP \vspace{1.5pt} $ &1.30 & 1.31& 3.68& 1.24&4.32 & 5.43& $>$ 6.77  \\ 
\hline
{  } & $\tau_{\rm eff, IU}$  & 6& 8& 20 & 20 & 200 & 1000 & $>$ 1600 \\
$\thetaB = 90^{\circ}$& $\tau_{\rm eff, AP}$ & 6& 6& 6 & 6 & 10 & 10 & 10 \\
{  }& $\timeIU / \timeAP \vspace{1.5pt} $&0.77 &1.16 & 4.58& 3.32 &112 &2183 & $>$ 3469  \\
\enddata
\end{deluxetable}
\vspace{-30pt}

The comparison of minimum slab effective optical depths \teq{\tau_{\rm eff, AP}} 
and \teq{\tau_{\rm eff, IU}} in Table~\ref{tab:IU_AP_inject_simulation}, which are
discretized to integer values, clearly indicates that higher \teq{\tau_{\rm eff}} are 
required for IU injections around or below the cyclotron frequency.  This is a 
consequence of the AP injection more closely resembling the photon anisotropy 
and polarization configuration at the same distance from the surface for 
\teq{\tau_{\rm eff}\to\infty} atmospheres.  In the super-cyclotronic domain, 
the injection protocol does not matter, generating similar \teq{\tau_{\rm eff}} 
and run times.  In fact, above around \teq{\omega/\wcyc \sim 1/\sqrt{3}}, the
{\sl MAGTHOMSCATT} code can be run with either injection choice with similar 
efficiency.  In contrast, when \teq{\omega/\wcyc} drops to around 0.1 or lower, 
the AP injection protocol quickly becomes more efficient and therefore preferred.  
This is particularly so for mid-latitude and equatorial locales (\teq{\thetaB \gtrsim 30^{\circ}}),
for which the AP choice leads to a much lower convergence \teq{\tau_{\rm eff}} and 
much shorter run times than IU does.  Notably, in a number of cases, the IU runs 
did not realize convergence for \teq{\omega/\wcyc=0.01} and locales away from 
the magnetic pole with simulation run times inferior to 24 hours on the cluster.  
Accordingly, the developments of precision high opacity AP configurations 
that are delivered in this paper are particularly helpful for the simulation of 
magnetar atmospheres, which generally sample \teq{\omega/\wcyc < 0.01} 
domains.  Again, the reason for this originates in the wide disparity of the 
scattering cross sections that are sampled over the full range of photon 
polarizations and propagation directions relative to \teq{\Bvec}. 

The high opacity radiative transfer analysis presented herein is useful 
for radiation dynamics considerations in various neutron star settings.  
The anisotropy elements addressed in Sec.~\ref{sec:anis_pol_char} enable
the specification of various radiation pressure tensor components in an anisotropic, 
magnetized plasma.  Knowledge of this pressure tensor is important for accurate 
modeling of super-Eddington environments.  The most notable of these are the common bursts 
and rare giant flares of magnetars, both of which have luminosities well above 
the Eddington limit, and via simple energetics considerations must be highly 
opaque to Thomson scattering \citep[e.g.,][]{Lin-2012-ApJ,Taverna-2017-MNRAS}.
Consequently, the intense radiation pressure in both these types 
of magnetar transients must drive the emitting plasma to move relativistically.  
The speeds of the motions are governed by dynamics, and will critically influence 
the spectrum of the radiation we detect \citep{Lin-2012-ApJ,Roberts-2021-Nature},
and likely its polarization also, thereby providing a science agenda for future hard X-ray polarimeters.
Accordingly, a deep understanding of radiation pressure in the burst/flare emission regions 
at various magnetospheric locales with different field strengths and plasma 
temperatures is required to forge detailed models of these short-lived events.
Such understanding is enabled by the specification of the high opacity forms for the 
scaled Stokes parameters in Eq.~(\ref{eq:hatIQV}) and resultant cross section 
in Eq.~(\ref{eq:sigma_tot_ave}) with great precision at a 
wide range of photon and electron cyclotron frequencies.   These forms can be 
readily employed in both cold and warm plasma settings in neutron star magnetospheres.  

The analysis can also be applied to the hydrostatics of accretion columns in high mass 
X-ray binaries, where near-Eddington luminosities suggest an active emission region 
that resides in the polar column at a location offset from the stellar surface due to intense 
radiation pressure \citep[e.g.][]{Becker-1998-ApJ,Schwarm-2017-AandA}, the 
historical ``fan-beam'' scenario.  The forms in Eq.~(\ref{eq:hatIQV}) in combination 
with the anisotropic cross section in Eq.~(\ref{eq:sigma_tot_ave}) can assist 
in improving the determination of the stand-off accretion shock altitude that was 
a focus in \cite{Becker-1998-ApJ}.  This in turn informs estimates of the neutron star
surface field that couple to the accretion shock fields obtained from hard X-ray cyclotron 
absorption lines in these systems.  While the {\sl MAGTHOMSCATT} simulation was 
envisaged and devised for neutron star surface applications, it is readily 
adaptable to high \teq{\taut} magnetospheric problems, both for X-ray binaries and
for magnetars.  The caveat with this is 
that ideally the cross section should be upgraded to the full QED magnetic Compton 
form \citep[e.g.,][]{DH-1986-ApJ,Gonthier-2014-PhRvD,Mushtukov-2016-PhRvD}, 
a somewhat formidable task given the mathematical complexity associated with 
the many cyclotron harmonics.  Efforts by our broader group to deliver such QED Compton
cross section forms that are convenient for radiative transfer simulations of magnetar 
emission zones are continuing (Gonthier et al., in preparation).  At the generally lower 
fields present in X-ray binary accretion columns, QED cross sections for radiative transfer 
near the cyclotron fundamental and low harmonics have been implemented in the 
simulations of \cite{Araya-1999-ApJ} and \cite{Schwarm-2017-AandA} in the quest 
for precision studies of cyclotron line formation. 

\section{Conclusion}
 \label{sec:conclude} 

For cold, magnetized plasmas, this paper has presented a radiative transfer equation analysis 
of the asymptotic configuration of radiation anisotropy and polarization in the limit of 
high Thomson scattering opacity, \teq{\taut = n_e\sigt h \gg 1}.  
The focus was on the influence of a strong magnetic field on the equilibrium photon population 
via the determination of Stokes parameters at different frequencies \teq{\omega}
relative to the cyclotron frequency \teq{\wcyc}.  The methodology distilled a phase matrix 
formalism down to two integral equations capturing information on the two eigenmodes 
of radiation propagation in the plasma.  These master equations constitute a Neumann problem 
that was solved numerically, additionally identifying useful analytic solutions
in the highly-magnetic, \teq{\omega \ll \wcyc}, and essentially non-magnetic,
\teq{\omega\gg \wcyc}, domains.  Based on these solutions and asymptotic 
analytic forms, empirical approximations for the two key parametric functions
\teq{{\cal A}(\omega )} (representing anisotropy) and \teq{{\cal C}(\omega )} (representing
circularity) were delivered.  These were then employed to illustrate how the magnetic 
Thomson Monte Carlo transport simulation  {\sl MAGTHOMSCATT} was made more efficient 
with the use of the high opacity anisotropy and polarization (AP) configuration in the 
photon injection protocol at the bottom of an atmospheric slab.  For different surface locales, 
Table~\ref{tab:IU_AP_inject_simulation} lists the values of the effective 
slab optical depth parameter \teq{\taueff} that guaranteed convergent determinations 
of the emergent anisotropy and polarization for photons exiting the top of the slab.
Code speed-up relative to isotropic/unpolarized injection protocols was also listed 
in Table~\ref{tab:IU_AP_inject_simulation}, and it was greatest at \teq{\omega\ll\wcyc} frequencies
and for magnetic fields lying closer to the atmospheric horizon (i.e., surface equatorial zones), 
a case germane also to the modeling of magnetar bursts in their magnetospheres.
The anisotropic pressure determinations for the high opacity photon configuration
will also prove useful in modeling the hydrostatics and dynamics of 
near-Eddington and super-Eddington neutron star environments such as 
accretion columns in X-ray binaries, and bursts and giant flares in magnetars.

\begin{acknowledgements}

We thank Alice Harding and also the anonymous referee for comments helpful to the polishing of the manuscript.
M.G.B. thanks NASA for generous support under awards 80NSSC22K0777, 80NSSC22K1576 and 80NSSC24K0589,
and the National Science Foundation for support via grant AST-1813649.
This work was supported in part by the Big-Data Private-Cloud Research Cyberinfrastructure 
MRI-award funded by NSF under grant CNS-1338099 and by Rice University's Center for Research Computing (CRC).

\end{acknowledgements}

\appendix

\centerline{Appendix A}
\vspace{2pt}
\centerline{Phase Matrix Elements}
\vspace{5pt}

Here we list the nine phase matrix elements that do not integrate
to zero in Eq.~(\ref{eq:redistrib_def}) on the azimuthal interval \teq{0 \leq \phi_{fi}  \leq 2\pi}.  
The other seven, being odd functions of \teq{\phi_{fi} } on \teq{[0,\, 2\pi]}, and therefore not 
germane to the radiative transfer developments of this paper, can be found in 
Eq.~(2) of \cite{Whitney-1991-ApJS} and pages 42-43 of \cite{Barchas-2017-thesis}.
For \teq{x=\omega /\wcyc}, using the scaled form \teq{\hat{P}_{if}= {P}_{if} \, \sigma (\omega,\, \mu_i )/r_0^2} 
identified in Eq.~(\ref{eq:Stokes_equil_matrix}), these elements are
\begin{eqnarray}
   \hat{P}_{11} & = & \sin^2\theta_i\sin^2\theta_f 
        + \dover{x^2}{(x^2-1)^2} \, \cos^2\theta_i\cos^2\theta_f 
             \Bigl[ x^2 \cos^2\phi_{fi}  + \sin^2\phi_{fi}  \Bigr] 
        + \dover{x^2}{2(x^2-1)} \sin 2\theta_i \sin 2\theta_f \, \cos \phi_{fi}   \nonumber\\[3.0pt]
   \hat{P}_{12} & = & \dover{x^2}{(x^2-1)^2} \, \cos^2\theta_f 
             \Bigl[ \cos^2\phi_{fi}  + x^2 \sin^2\phi_{fi}  \Bigr]  \nonumber\\[3.0pt]
   \hat{P}_{14} & = & \dover{x^3}{(x^2-1)^2} \cos\theta_i \cos^2\theta_f
              + \dover{x}{2(x^2-1)}\, \sin \theta_i \sin 2\theta_f \, \cos \phi_{fi}  \nonumber\\[3.0pt]
   \hat{P}_{21} & = & \dover{x^2}{(x^2-1)^2} \, \cos^2\theta_i 
             \Bigl[ \cos^2\phi_{fi}  + x^2 \sin^2\phi_{fi}  \Bigr]   \nonumber\\[3.0pt]
   \hat{P}_{22} & = &  \dover{x^2}{(x^2-1)^2} \, \Bigl[ x^2 \cos^2\phi_{fi}  + \sin^2\phi_{fi}  \Bigr]    
 \label{eq:calM_comps} \\[3.0pt]
   \hat{P}_{24} & = & \dover{x^3}{(x^2-1)^2} \, \cos\theta_i \nonumber\\[5.0pt]
   \hat{P}_{41} & = & \dover{2 x^3}{(x^2-1)^2} \, \cos^2\theta_i\cos\theta_f 
        + \dover{x}{x^2-1} \sin 2\theta_i \sin \theta_f \, \cos \phi_{fi}   \nonumber\\[3.0pt]
   \hat{P}_{42} & = & \dover{2x^3}{(x^2-1)^2} \, \cos\theta_f \nonumber\\[5.0pt]
   \hat{P}_{44} & = & \dover{x^2 (x^2+1)}{(x^2-1)^2} \cos\theta_i \cos\theta_f
              + \dover{x^2}{x^2-1}\, \sin \theta_i \sin \theta_f \, \cos \phi_{fi}  \quad .\nonumber
\end{eqnarray}
These are the forms that are inserted into the re-distribution function integral 
in Eq.~(\ref{eq:redistrib_def}).  Observe that the last terms of \teq{\hat{P}_{11}}, \teq{\hat{P}_{14}}, 
\teq{\hat{P}_{41}} and \teq{\hat{P}_{44}} all do not contribute to \teq{{\cal R} (\mu_i,  \mu_f )} after 
the \teq{\phi_{fi} } integration under our azimuthally symmetric assumption.  Note
that a factor of \teq{1/2} appears in the third term of  \teq{\hat{P}_{11}}, correcting 
a typographical error in \cite{Whitney-1991-ApJS} that was identified by \cite{Barchas-2017-thesis}.

Note also that we retain the azimuthal phase convention of \cite{Whitney-1991-ApJS} that
differs from that of \cite{Chou-1986-ApSS} and \cite{Barchas-2017-thesis}, wherein 
an extra minus sign appears in the various Stokes \teq{V} terms in  \teq{\hat{P}_{14}}, 
\teq{\hat{P}_{24}}, \teq{\hat{P}_{41}} and \teq{\hat{P}_{42}}.  The origin of this sign is in the form
chosen for specifying a photon's complex electric field vector.  Here we adopt
the choice in Eq.~(1) of \cite{Barchas-2021-MNRAS}, wherein \teq{\mathbf{E}(t) \propto 
e^{-i\omega t}}, contrasting the choice of \teq{\mathbf{E}(t) \propto e^{+i\omega t}} in 
\cite{Chou-1986-ApSS}.  Thus switching the sign of \teq{\omega} merely amounts to 
changing that for \teq{V} throughout.

\newpage


\bibliographystyle{aasjournal}
\bibliography{apj25bhd_magThom_biblio}

\end{document}